\documentclass{aa}
\usepackage{graphicx}
\usepackage{txfonts}
\usepackage{xcolor}

\usepackage[colorlinks=true, linktocpage, linkcolor={blue!60!black}, citecolor={blue!60!black}, urlcolor={blue!60!black}]{hyperref}
\begin{document}

  \title{Radio-relic and the diffuse emission trail discovered in a low mass galaxy cluster Abell 1697}
\titlerunning{Discovery of relic in Abell 1697}
\authorrunning{Paul et al.,}

   \author{Surajit Paul
          \inst{1}\fnmsep\thanks{Email:surajit@physics.unipune.ac.in}, 
        Sameer Salunkhe\inst{1},   
        Satish Sonkamble\inst{2}, 
        Prateek Gupta\inst{1},
        Tony Mroczkowski\inst{3},
        and
        Somak Raychaudhury\inst{4}}
               
        \institute{Department of Physics, Savitribai Phule Pune University, Pune 411007, India; \email{surajit@physics.unipune.ac.in}
         \and
             National Centre for Radio Astrophysics, Tata Institute of Fundamental Research, Post Bag 3, S. P. Pune University Campus, Ganeshkhind, Pune 411007, Maharashtra, India
             \and
             European Southern Observatory (ESO), Karl-Schwarzschild-Str. 2, D-85748 Garching, Germany
             \and
             Inter University Centre for Astronomy and Astrophysics, Pune 411007, India
             }

   \date{Received September 15, 1996; accepted March 16, 1997}

 
  \abstract
{We report the discovery of a putative radio relic, 830 kpc in length and found toward the outskirts of galaxy cluster Abell 1697 ($z=0.181$), using the LOFAR Two Meter Sky Survey (LoTSS) at 144 MHz. With an X-ray-inferred mass of $M_{500}^{X-ray}=2.9^{+0.8}_{-0.7}\times10^{14}~\rm{M_{\odot}}$, this places Abell 1697 among the least massive relic hosts. The relic is also detected at 325 MHz in the Westerbork Northern Sky Survey (WENSS) and at 1.4 GHz in the NRAO VLA Sky Survey (NVSS) with an average spectral index of $\alpha(144,325,1400~\rm{MHz})=-0.98\pm0.01$ and magnetic field of $B_{eq}\sim0.6~\mu$G. This relic, located in the northeast periphery of the cluster, is 300 kpc wide, exhibits a gradual spectral steepening across the width ($\alpha_{144 \rm{MHz}}^{1.4\rm{GHz}}(inj)=-0.70\pm0.11$ to $\alpha_{144 \rm{MHz}}^{1.4\rm{GHz}}(edge)=-1.19\pm0.15$), as well as indications of a co-spatial X-ray (ROSAT) shock and the radio relic emission. The radio power of the relic is $P_{1.4GHz}=8.5\pm1.1\times10^{23}~\rm{W\;Hz^{-1}}$, which is found to be in good agreement with the expected empirical correlation between the radio power and Largest Linear Size (LLS) of relics. The relic is trailed by extended ($790\times550$ kpc) diffuse radio emission towards the cluster center, that is likely an ultra-steep spectrum ($\alpha_{144 \rm{MHz}}^{1.4\rm{GHz}}<-1.84$) radio source. This structure is also found to be older by at least 190 Myrs, has a very low surface brightness of $0.3~\mu$Jy arcsec$^{-2}$ and magnetic field $B_{eq}\sim0.8~\mu$G, similar to that of a radio phoenix. Finally, we discuss the possible mechanisms responsible for the relic and the trailing diffuse radio emission, invoking re-acceleration due to wake turbulence, as well as the revival of fossil electrons from an old AGN activity by the cluster merger shocks.}


   \maketitle
%

\section{Introduction}\label{intro}

With the improved sensitivity of radio telescopes in recent years, diffuse and elongated radio sources---usually known as radio relics---are observed in a growing number of galaxy clusters \citep[c.f.\ review by][]{Weeren_2019SSRv}. The largest among these ($\gtrapprox$1 Mpc) are the bow-shock relics, which occur at the peripheries of galaxy clusters. They are low surface brightness $\sim$ 0.1 - 1 $\mu$Jy arcsec$^{-2}$ at 1.4 GHz, highly polarised (20-30\%), and are thought to originate from non-thermal synchrotron emission \citep[e.g.][]{Weeren_2019SSRv}. These spectacular objects can be of two major types, symmetric doubles \citep[e.g. Abell 3376,][]{Bagchi_2006Sci}), or single bow-like relics \citep[e.g. Abell 2744,][]{Paul_2019MNRAS} termed as `cluster relic shocks' \citep{Weeren_2019SSRv} because of the strong evidence in favour of their association with cluster merger shocks.

Galaxy cluster mergers are the most energetic events in the Universe since the Big Bang itself, releasing a tremendous amount ($\sim10^{64}$ ergs) of energy \citep{Sarazin_2002ASSL} within a time span of $\gtrsim1$ Gyr. This creates an extreme pressure gradient in the intracluster medium (ICM), causing expansion at supersonic speeds \citep{Paul_2012JPhCS}, and producing strong Mpc-scale shocks that travel across the cluster medium, reaching the cluster periphery. In the process, they dissipate energy by heating and generating turbulence in the ICM \citep{Kang_1996ApJ,Paul_2011ApJ}.  These collision-less shocks are also known for efficiently accelerating charged particles via diffusive shock acceleration (DSA); \citealt{Drury_1983RPPh}) and magnetic field amplifiers \citep{Iapichino_2012MNRAS}, eventually producers of radio emission through synchrotron process, creating radio relics \citep{Kang_2012ApJ}. Alternatively, merger shocks can also re-accelerate fossil or ghost electrons from the past events of radio galaxies through compression, to produce relic emissions called the phoenix \citep{Ensslin_2001A&A,Kang_2012ApJ}. 

Relativistic electrons loose energy very fast through synchrotron emission and inverse Compton cooling with an average life of about $10^{6}$ years, indicating a surviving structure of just about 100 kpc \citep{Kang_2017ApJ}. But, the radio structures in discussion, are of Mpc in size that remains visible for Gyrs. Therefore, to keep alive a radio structure of an order larger in size and timescale, the constant injection of relativistic electrons is absolutely necessary. Cluster merger shocks are the in-situ particle accelerators, reputedly injecting relativistic electrons continuously \citep{Ensslin_1998A&A}, keeping the bow-shock radio relics alive for long ($>$ Gyr). These shocks are analogous to the tsunami waves, becoming stronger (increasing in Mach number) as they reach the cluster periphery \citep{Ensslin_Science2006,Paul_2011ApJ}, efficiently accelerating charged particles at the cluster outskirts through DSA mechanism, making radio emission better visible as cluster peripheral relics. Since particle acceleration happens only at a small width near to the shock front, the spectral index of peripheral radio emission steepens behind the surface of injection due to synchrotron losses or ageing. Therefore, the spectral index distribution observed across these relics, typically show a flat spectrum at the outer edge that gradually steepens towards the inner edge \citep{Stroe_2014MNRAS}. Radio phoenixes, by contrast, are due to re-energizing of the aged electrons from the Mpc scale old AGN lobes by low Mach shocks \citep{Gasperin 2015MNRAS}, thus producing relic radio emission with extremely steep spectrum with curvature. They can be found both in the central region and the outskirts of the clusters \citep{Hoeft_2004MNRAS}.

Relics are detected more commonly in massive merging systems \citep{Gasperin_2014MNRAS}. Low mass clusters and groups are however, more affected by mergers and various non-gravitational processes \citep{Lovisari_2015A&A}, and contain high levels of cosmic-rays \citep{Jubelgas_2008A&A} likely due to frequent AGN feedback events \citep{Gilmour_2007MNRAS,Sivakoff_2008ApJ}. Since the sound speed in the ICM depends on ICM temperature, a merger of given velocity can produce a much higher Mach number shock in deeper core ICM of low mass, low temperature clusters than in higher mass clusters \citep{Sarazin_2002ASSL}. This would mean, if all relics are shock generated, their fraction may be observed to increase in low mass systems as the sensitivity of radio and X-ray surveys improve. But, so far, only a few relics have been reported to be found in low mass clusters \citep{Gasperin_2014MNRAS,Gasperin_2017A&A,Kale_2017MNRAS,Dwarakanath_2018MNRAS}.

We made the serendipitous discovery of a large diffuse radio relic emission in the low mass ($M_{500}<5\times10^{14}\;\rm{M_{\odot}}$) cluster Abell 1697 while searching the field of cluster Abell 1682 in the LOFAR Two Meter Sky Survey (LoTSS). With further inspection, the relic-like emission in the cluster outskirts is also detected at 325 MHz WENSS and at high frequency 1.4 GHz NVSS maps. Here, in this work, we present a multi-wavelength study using the radio LoTSS, WENSS, NVSS, TGSS-ADR and VLA First survey, ROSAT survey in X-ray and Pan-STARRS1 optical survey data to report this discovery. We introduce the paper in Section~\ref{intro}. Detailed multi-wavelength properties of the cluster is presented in Section~\ref{object}. We discuss our findings and conclude the paper in Section~\ref{discus} and Section~\ref{conc} respectively.

We assumed $\Lambda$CDM cosmology with parameters $H_0=70.2$, $\Omega_M=0.3$, $\Omega_{\Lambda}=0.7$ for this study. The redshift of the cluster is $z=0.181$, yielding a luminosity distance $D_L=869.2$ Mpc and angular scale of 3.026 kpc per arcsecond.

\section{Cluster Abell 1697}\label{object}

Abell 1697, also known as RXC J1313.1+4616, is a cluster in the northern sky at 13:13:04.56, +46:15:52.7 (J2000). This moderately distant cluster at a spectroscopic redshift of $z=0.1813\pm0.0006$ \citep{Clerc_2016MNRAS} hosts 84 galaxies within a radius of 9 arcmin \citep{Gal-Yam_2008ApJ}. Its colour dependent stellar mass is reported to be $1.32\pm0.31\;\times 10^{13}\;M_{\odot}$ \citep{Sharon_2007ApJ}, while the total SZ mass is $M^{SZ}_{500}=4.34^{+0.32}_{-0.33}\;\times10^{14}\;\rm{M_{\odot}}$ ({\it Planck} Catalogue: PSZ2 G111.75+70.37, \citealt{Planck_2016A&A}). The cluster has X-ray brightness of $1.5\pm0.3\times 10^{44} \rm{erg\;s^{-1}}$ (0.1-2.4 keV, ROSAT survey, \citealt{Clerc_2016MNRAS}). While, no X-ray estimated mass is reported yet, we provide estimates based on luminosity-mass scaling relations in section~\ref{x-ray}.

\subsection{Radio signatures}\label{radio-sig}

\begin{figure}
\vspace{0.2cm}
\includegraphics[width=9cm]{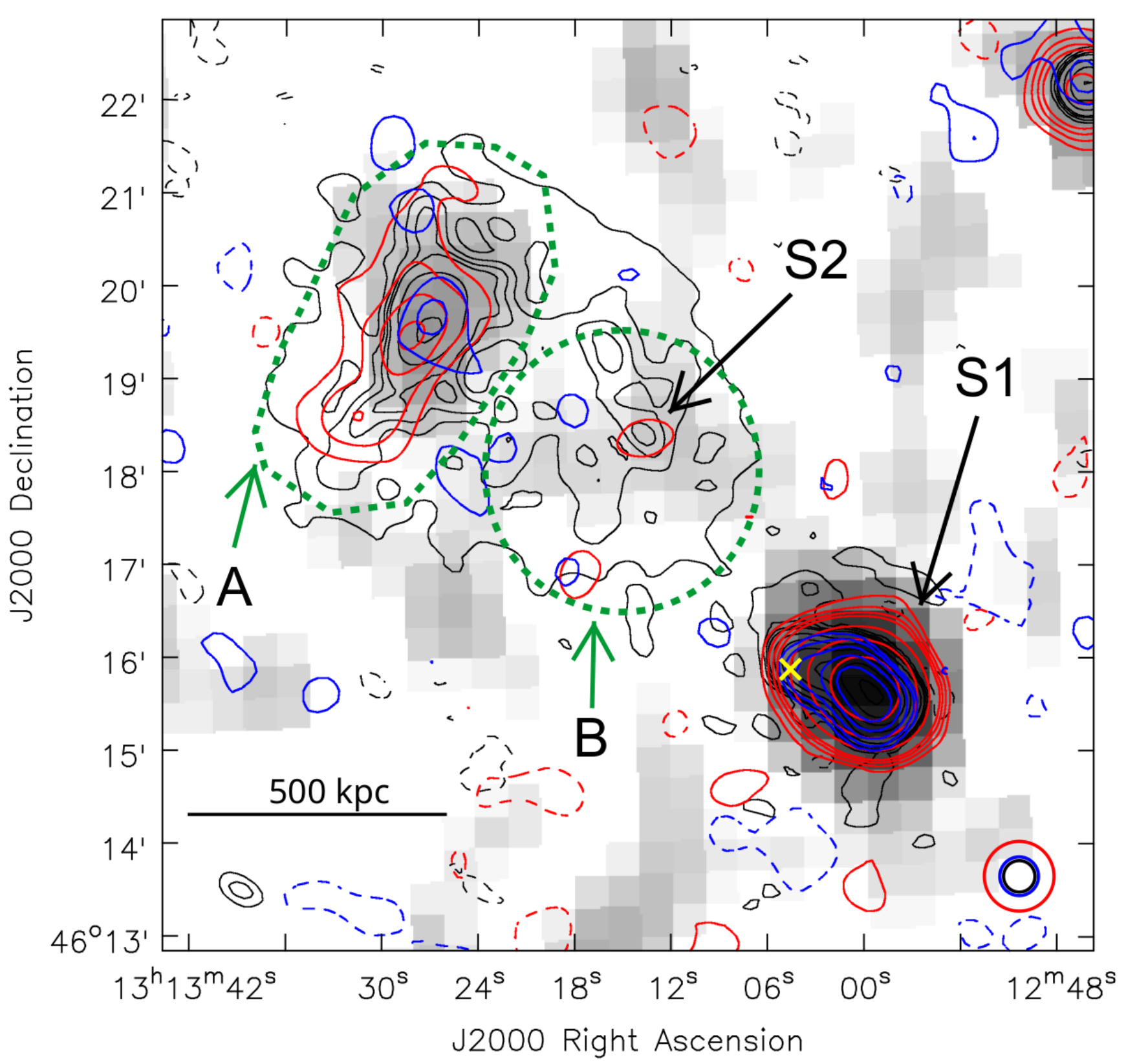}
\caption{Contours of low resolution ($20^{\prime\prime}\times20^{\prime\prime}$) LoTSS at 144~MHz (black), NVSS at 1.4~GHz (red) and TGSS at 150~MHz (blue) on WENSS (325 MHz) gray colour map. Contours are at  $-3,5,9,13,17,34,68,136,272,544,1088 \times\sigma$ for LoTSS with rms $\sigma=150~\mu$Jy, NVSS at $-3,3,5,7,9,18,36,72,144 \times\sigma$ with $\sigma=350~\mu$Jy and TGSS at $-3,3,5,10,20,40,80,160 \times\sigma$ with $\sigma=3$~mJy. The relic and the trailing radio emission is highlighted by green dashed polygon (`A') and a circle (`B') respectively. `S1' and `S2' are two point like radio sources inside the cluster. The Abell cluster centre is marked as yellow cross.}\label{fig:radio-lofar}
\end{figure}

\begin{figure}
\includegraphics[width=9cm]{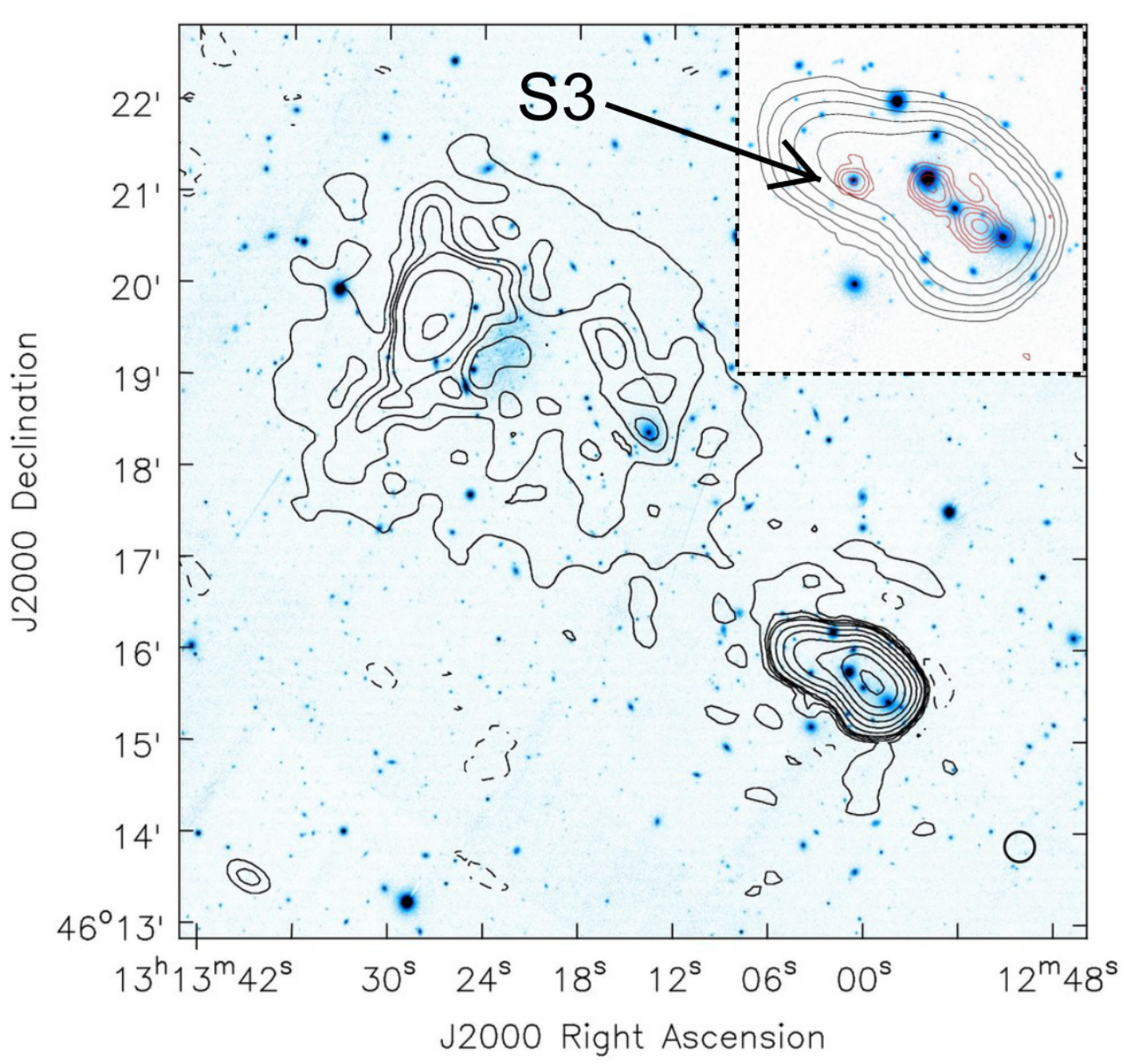}
\caption{Pan-STARRS1 `i' band optical image overplotted with LoTSS low-resolution contours (same as Fig.\ref{fig:radio-lofar}). {\bf Inset:} High-resolution VLA First and LOFAR contours for the central radio source are overplotted on Pan-STARRS1 image. VLA First contours are at $3,6,12,24,48 \times150~\mu$Jy and LoTSS contours are at $10,20,40,80,160 \times\sigma$. `S3' (13:13:03.0, 46.15.48.6) is the  third radio point source in Abell 1697.} \label{fig:optical-overlay}
\end{figure}

\begin{table*}
\caption{(a) Names of the surveys, their respective rms and beam size is given in the first three columns. Size, flux density and power of the proposed relic in the next three columns. Final two columns reads the size and flux density of the trailing diffuse emission. (b) Position, flux and radio powers of each of the point sources (i.e. S1 and S2) are given in columns $2^{nd}$ to $7^{th}$ respectively.}
\label{tab:imdtls}      
\centering
 \begin{tabular*}{\textwidth}{@{\extracolsep{\fill}}l c c c c c c c c} 
 \hline
 \multicolumn{3}{l}{(a) Diffuse radio structures}\\
 \hline
Survey & rms & beam& \multicolumn{3}{c}{Relic} & \multicolumn{2}{c}{Trail (phoenix?)}\\
(Frequency (MHz))  & ($\mu$Jy beam$^{-1}$) & ($^{\prime\prime}$) & [Size ($^{\prime\prime}$) & Flux (mJy) & Power (W Hz$^{-1}$)] & [Size ($^{\prime\prime}$) & Flux (mJy) ]\\
 \hline\hline
LOFAR (144)& 150 & $20\times20$, PA 90$^{\circ}$ & $275\times100$ & $97.1\pm9.8$ &  $8.8\pm0.9\times10^{24}$&$260\times180$ & $96.9\pm9.8$\\ 
 TGSS-ADR1 (150)  & 3000 & $25\times25$, PA 0$^{\circ}$ & -- & -- &  --&--&--\\ 
WENSS (325)& 2000 & $54\times54$, PA 0$^{\circ}$&$142\times85$ &$34.9\pm3.5$ & $3.2\pm0.3\times10^{24}$ & -- & --\\
NVSS (1.4 GHz)  & 350 & $45\times45$, PA 0$^{\circ}$ & $205\times75$ & $9.4\pm1.2$ &  $8.5\pm1.1\times10^{23}$&--&--\\
\hline\hline\\
\hline
  \multicolumn{3}{l}{(b) Radio point sources}\\
 \hline
Survey & \multicolumn{3}{c}{Source `S1'} & \multicolumn{3}{c}{Source `S2'}\\
 &\multicolumn{2}{c}{[Position} & Flux (mJy)] & \multicolumn{2}{c}{[Position} & Flux (mJy)] & \\
 & (RA: & DEC:)& & (RA: & DEC:)& &\\
 \hline\hline
 
LOFAR & 13:12:59.6 &               46.15.40.3 & 449.0$\pm$44.9 & 13:13:13.7 &              46.18.25.0 & 3.6$\pm$0.4&  \\ 
TGSS-ADR1 & & & 413.2$\pm$42.0 & & & -- &\\
WENSS & & &307.8$\pm$31.2 & & &--&\\
NVSS & & & 51.1$\pm$5.2 & & & 0.4$\pm$0.2\\
 \hline \hline
 \end{tabular*}
\end{table*}

In the field of Abell 1697, no diffuse radio relic or halo emission was previously reported. We report the newly-discovered extended, diffuse radio emission using LoTSS, Data Release 1 \citep{Shimwell_2019A&A}. LoTSS provides a very deep image with $150\;\mu$Jy~Beam$^{-1}$ at 144 MHz. The black contour map in Figure~\ref{fig:radio-lofar} shows the diffuse emission above 5$\sigma$. The region demarcated as `A' resembles peripheral relic structure. The proposed relic is also observed at 325 MHz WENSS (Gray) and at 1.4 GHz NVSS (red contour), and only the brightest part is visible in TGSS-ADR1 at 150~MHz (blue contour). This relic is also attached to an extended trailing diffuse radio emission (as seen in LoTSS contour) towards the cluster centre, marked as `B'.

In Figure~\ref{fig:optical-overlay},  we overlaid radio contours of LoTSS on PanSTARRS1 optical image which shows that the extended radio emission on the North-East (NE) side of the cluster cannot be linked to any point sources, confirming its diffuse nature. The central radio source (`S1') that appears to be a point-like in NVSS, WENSS and extended one in TGSS as well as in LoTSS maps, resolved in VLA FIRST survey (see red contours in Fig~\ref{fig:optical-overlay}, inset). Apparently, it consists of a radio double and a radio galaxy towards East of the double source. These radio point sources are found to be well correlated with their optical counterparts. A diffuse optical emission can also be noticed just behind the relic front. A further investigation with GALEX Near and Far ultraviolet (UV) images and reported HII emission \citep{Strobel_1991ApJ} indicate a possible star-forming region. This region is far behind the peripheral radio relic structure, ruling out having any physical connection between these two emissions.

We computed the flux densities ($S$) of observed radio sources within 3$\sigma$ contours ($\sigma$ is noise rms of the images given in Table~\ref{tab:imdtls}, 5$\sigma$ for LoTSS) and flux density errors ($\sigma_S$) using the usual relation
\begin{equation}\label{eq:Flux_err}
\sigma_S = \sqrt{(0.1 S)^2 + N ({\sigma})^2}
\end{equation}
where N is the number of beams covered by the total diffuse emission and $0.1S$ (i.e., 10\% of $S$) was assumed as the possible error due to calibration uncertainties.

The peripheral relic towards NE of the cluster is a bright source at 144 MHz with flux density of $97.1\pm9.8$~mJy and size of $830\times300$~kpc. The relic visible in NVSS map is of $620\times225$~kpc. It is a low brightness object at 1.4 GHz with a flux density of $9.4\pm1.2$ mJy and radio power of $8.5\pm1.1\times10^{23}$~W~Hz$^{-1}$. The large trailing emission behind the shock observed in LoTSS map (`B' in Fig~\ref{fig:radio-lofar}) is about 790~kpc long and 550~kpc in width having flux density of $96.9\pm9.8$~mJy (at 144 MHz). Except for a bright point source (`S2'), the trailing emission is not well detected at NVSS or WENSS map and nothing is observed in TGSS map. The point-like source (`S1' in Fig.~\ref{fig:radio-lofar}) observed in all the survey's is a very bright source at low frequency (144 MHz) with a flux density of $449.0\pm44.9$ mJy. Further, detailed parameters measured at different observed frequencies are given in Table~\ref{tab:imdtls}~(a)\&(b).

\subsubsection{Radio spectral properties}

\begin{figure*}
\includegraphics[width=9cm]{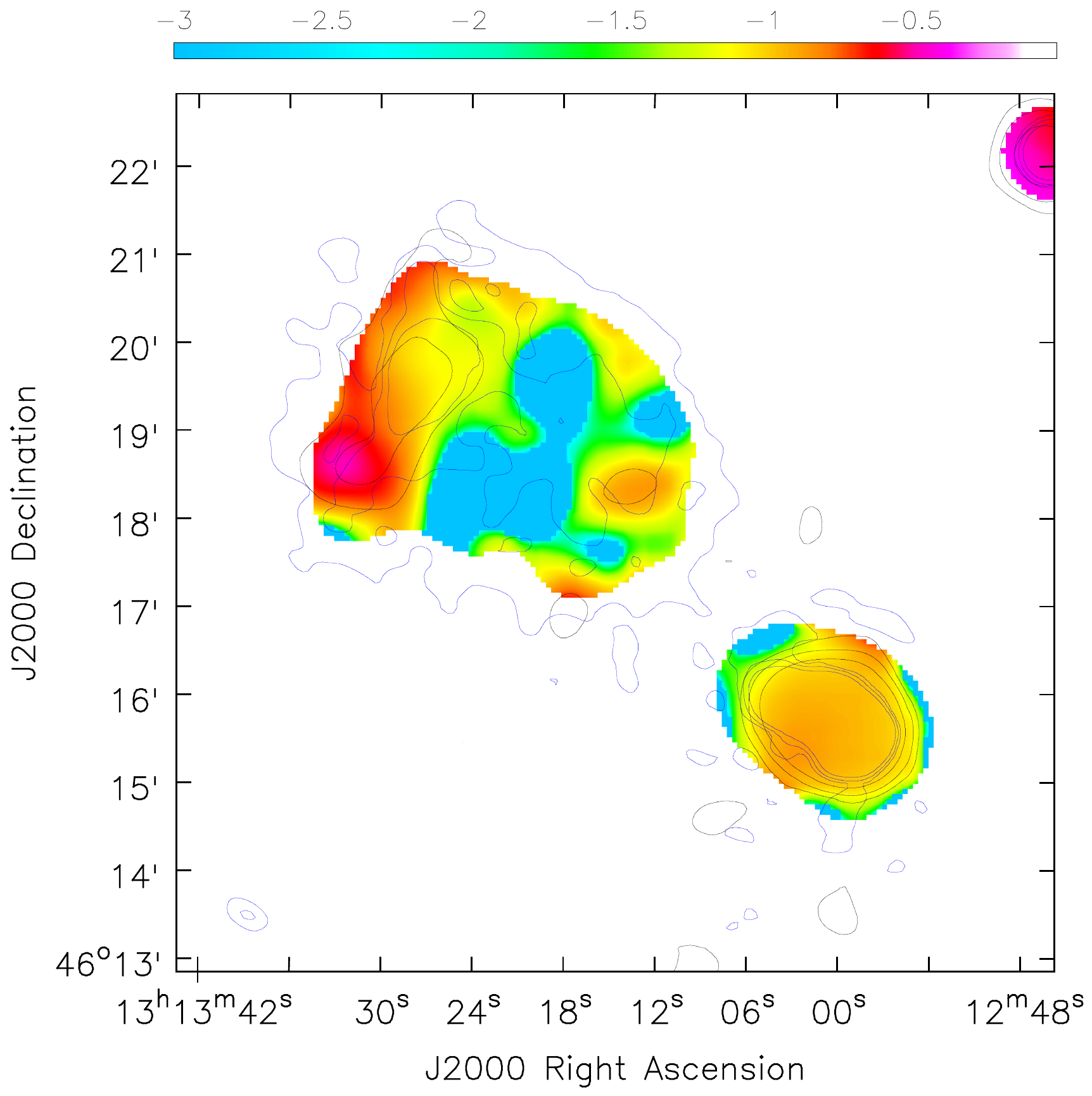}
\includegraphics[width=9cm]{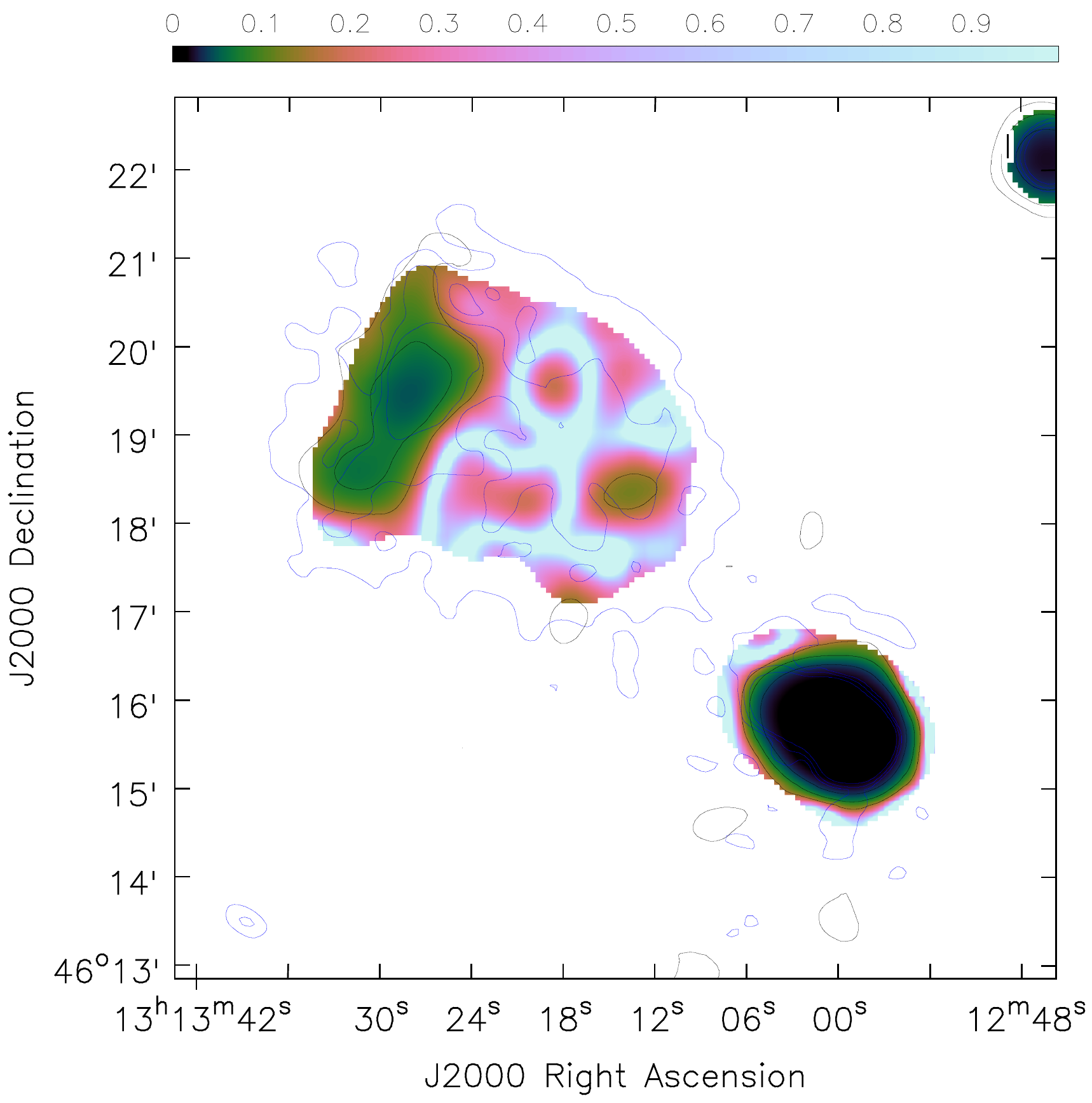}\\
\caption{{\bf Left panel:} Spectral index map using LOFAR 144 MHz and NVSS 1.4 GHz. {\bf Right panel:} Spectral index error map for the same}\label{fig:radio-spix}
\end{figure*}

The images were first re-gridded with IMREGRID task of CASA and thereafter beams are convolved to the beam of 1.4 GHz NVSS  i.e.\ $45^{\prime\prime}\times45^{\prime\prime}$ with the IMSMOOTH task. The UV coverage cannot be homogenized as these are survey images from different telescopes. The images were then masked below $3\sigma$ using the NVSS image. Finally, we computed the spectral index using CASA IMMATH task with the relation,
\begin{equation}\label{eq:Sp-Ind}
\alpha = \frac{\log(S_{\nu_2}/S_{\nu_1})}{\log(\nu_2/\nu_1)}
\end{equation}
where $S_{\nu_\#}$ and $\nu_\#$ (with $\nu_1 > \nu_2$ ) are the flux densities and the respective observed frequencies. Spectral index error map is made using the relation given below \citep{Kim_2014JKAS}
\begin{equation}\label{eq:SP-error}
\alpha_{err}(\alpha_{\nu_2,\nu_1})=\frac{1}{\log(\nu_2/\nu_1)}\times\left[\frac{\sigma_{\nu_1}^2}{I_{\nu_1}^2}+\frac{\sigma_{\nu_2}^2}{I_{\nu_2}^2}\right]^{\frac{1}{2}}
\end{equation}
with $I$ as the total intensity at respective frequencies at each of the pixels.

Figure~\ref{fig:radio-spix}~left panel shows a spectral index map of the studied radio features combining the images at 144~MHz (LoTSS) and 1.4~GHz (NVSS). The corresponding error map is plotted in Figure~\ref{fig:radio-spix}~right panel. The average spectral index of the proposed relic part (i.e. `A', with 3$\sigma$ detection of NVSS) is about $\alpha=-0.96\pm0.17$. The spectral index at the shock front (i.e., at injection) is $\alpha_{inj}=-0.70\pm0.14$, which steepens across the relic gradually and falls to $\alpha_{edge}=-1.19\pm0.23$. The average combined spectral index of the relic is found to be $\alpha(144,325,1400~\rm{MHz})=-0.98\pm0.01$. Because of non-detection of the structure `B' in both NVSS and WENSS, we computed the upper limits using the local rms of 0.35~mJy~beam$^{-1}$ (NVSS) and 2~mJy~beam$^{-1}$ (WENSS) respectively, and excluded the point source, marked as `S2' in Figure~\ref{fig:radio-lofar}. Upper limits for the spectral index thus computed are $\alpha_{144 MHz}^{1.4 GHz} (trail)<-1.84$ \& $\alpha_{144 MHz}^{330 MHz} (trail)<-2.02$. The average spectral index for the source `S1' is computed as $\alpha_{144\rm{MHz}}^{1.4\rm{GHz}}=-0.96\pm0.02$. Using the relation of Mach number and the spectral index i.e. 
\begin{equation}\label{eq:Mach-comp}
\mathcal{M}^2 = \frac{2\alpha-3}{2\alpha+1}
\end{equation} 
(considering DSA mechanism; see \citealt{Blandford_1987,Colafrancesco_2017MNRAS}), we computed the average Mach number of the shock as $\mathcal{M}=2.31^{+0.50}_{-0.27}$ (using the average spectral index of $\alpha_{144\rm{MHz}}^{1.4\rm{GHz}}=-0.96\pm{0.17}$) for the proposed relic.

\subsection{X-ray properties}\label{x-ray}

\begin{figure}
\includegraphics[width=9cm]{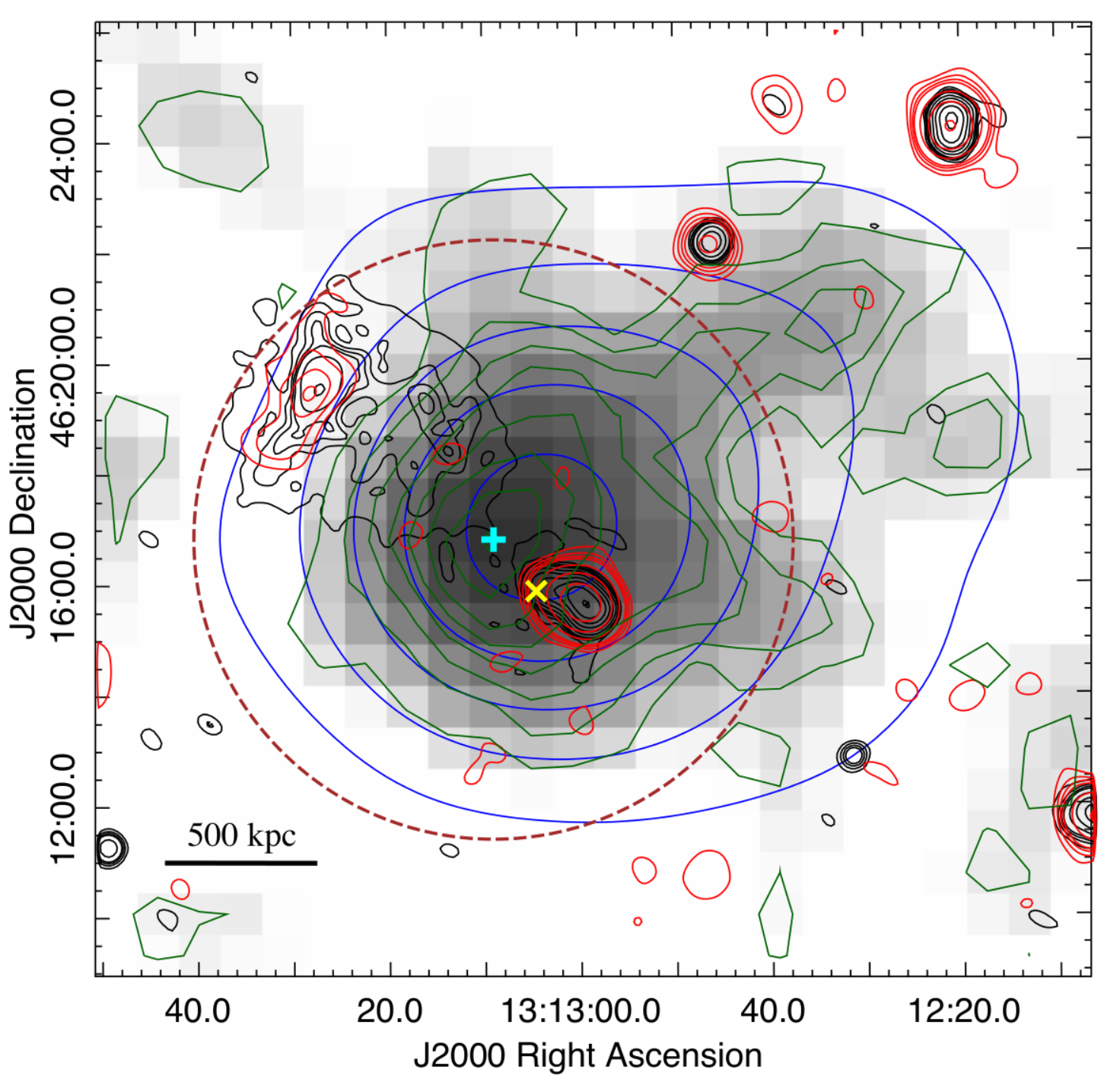}
\caption{ROSAT X-ray photon count map (0.1-2.4 keV) presented in gray, mildly and highly Gaussian smoothed contours are in dark green and blue respectively. NVSS and LoTSS radio contours are shown in red and black respectively. Red dashed circle represents $r_{500}$ radius, where the X-ray peak (centre) and the Abell cluster centre were marked as cyan `+' and yellow `$\times$' respectively.}\label{fig:x-ray} 
\end{figure}

Figure~\ref{fig:x-ray} shows the photon count map of ROSAT survey at 0.1-2.4 keV. Gray colour represents the photon counts and dark green contours are the mild Gaussian smoothed map. NVSS Radio flux is over-plotted as red contours. The ROSAT X-ray survey data is of just 0.6 ks, inadequate to create any X-ray brightness or temperature maps. Nonetheless, the dark green contours of the mildly smoothed photon counts show an overall asymmetric morphology, clumpy and extended towards North-West. We also find that the photon count does not vary smoothly or decline monotonically.

\begin{figure*}
\includegraphics[width=9cm]{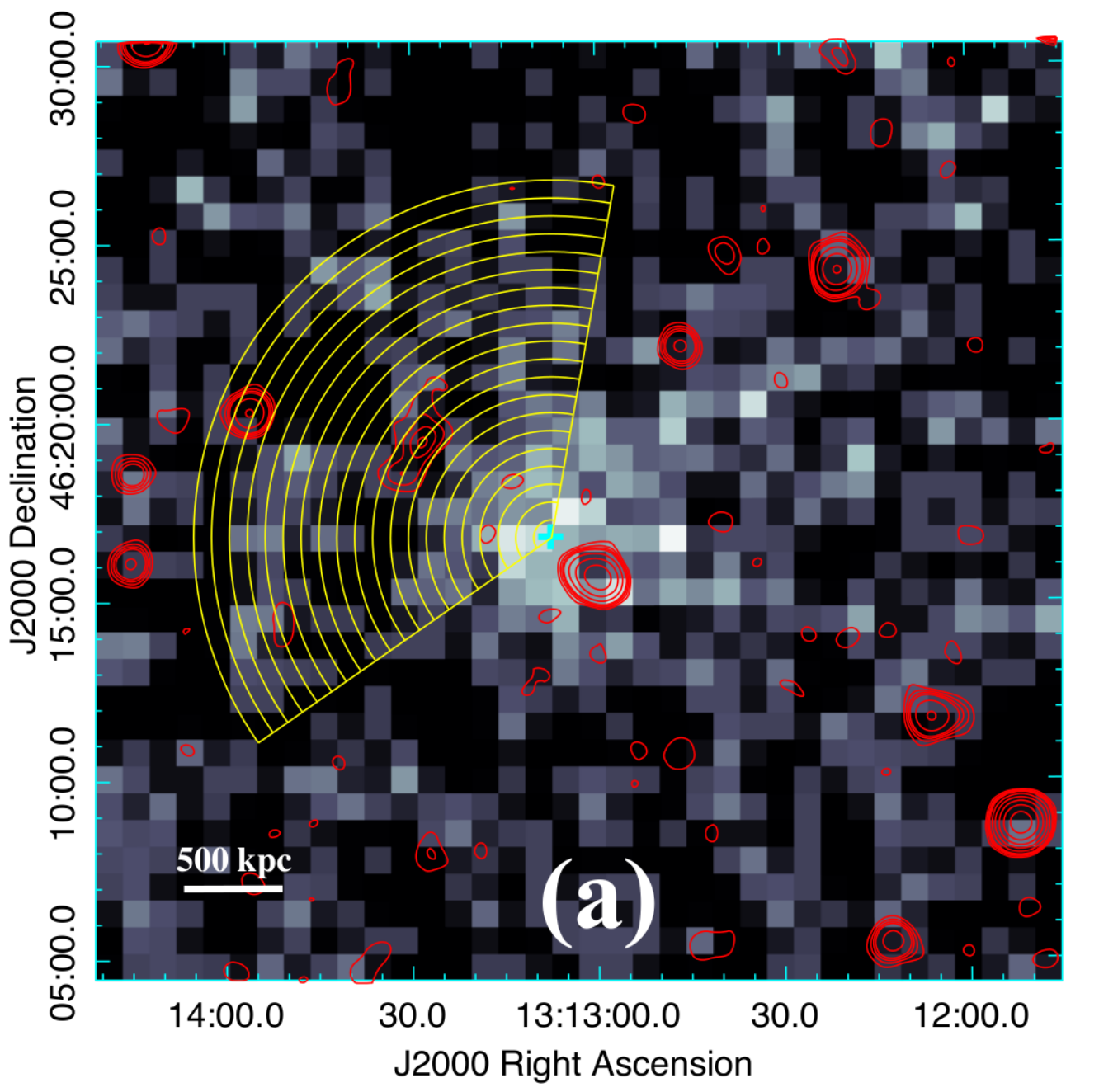}
\includegraphics[width=9cm]{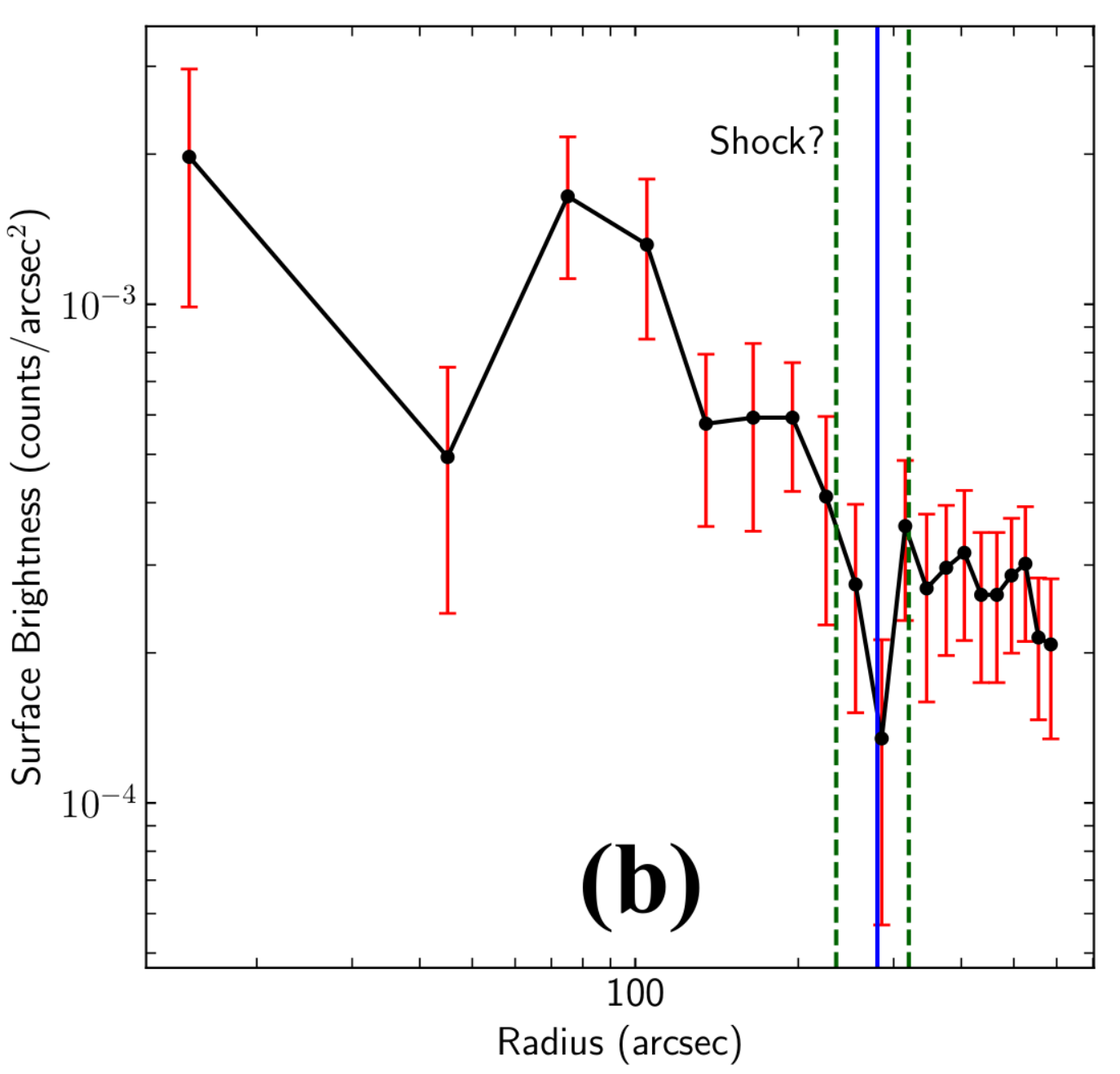}\hspace{-0.2cm}\\
\caption{{\bf Panel a:} Shows the sector for which radial profile of X-ray photon count (0.1-2.4 keV ROSAT survey data) is displayed in panel~(b). {\bf Panel b:} Photon count radial profile for the sector containing the radio relic shown in panel~(a) and the cluster radio shock position (blue vertical line). Distance of the inner and the outer edge of the radio relic (from the X-ray peak) are indicated with dark green vertical dashed lines.}\label{fig:x-ray1} 
\end{figure*}

The observed peak and extreme end of the radio relic-like diffuse emission (see red NVSS contours towards NE) are about $270^{\prime\prime}$ and $340^{\prime\prime}$ (i.e., about 815 kpc and 1.02 Mpc respectively) away from the X-ray peak or the X-ray centre (peak of mild Gaussian smoothed map at 13:13:08.6, +46:16:48.9, marked as cyan plus in Fig.~\ref{fig:x-ray}). Further, we plotted the radial profile of photon count for a sector that encloses the radio relic in Figure~\ref{fig:x-ray1}(b). A dip in the profile, possibly indicating a shock front at around $280\pm23^{\prime\prime}$ from the X-ray centre (i.e. $850\pm70$ kpc from the Gaussian peak fit). This feature overlaps with the cluster radio shock (relic) and the shock front as shown in Figure~\ref{fig:x-ray1}(b) with green vertical dashed-lines (at $235^{\prime\prime}$ and $320^{\prime\prime}$ i.e. $\sim710-970$~kpc). We caution the reader that ROSAT data have a large point spread function and low significance, making shock identification exceedingly difficult at moderate redshifts such as this one.

The reported X-ray luminosity for photon energy range of 0.1-2.4 keV (ROSAT) is $L_{500}=1.5\pm0.3\times10^{44}$~erg~s$^{-1}$ (SPIDERS clusters, \cite{Clerc_2016MNRAS}). Adopting the method described by \citet{Mantz_2018MNRAS}, equation~no.~6, we have estimated a mass of $2.9^{+0.8}_{-0.7}\times10^{14}\;\rm{M_{\odot}}$. The reported SZ mass of $M^{SZ}_{500}=4.34^{+032}_{-0.33}\times10^{14}\;\rm{M_{\odot}}$ is in rough agreement with the mass computed from X-ray data.
Using the virial radius relation derived from the self-similarity \citep{kaiser_1986MNRAS} i.e., 
\begin{equation}\label{eq:Mass-rad}
r_{vir} = \left[3M_{200}/(4\pi\nabla_{c}(z)\rho_{cr}(z))\right]^{1/3}
\end{equation}
(where $M_{200}$ is virial mass, $\nabla_{c}(z)$ and $\rho_{cr}(z)$ are the over density ratio and critical density at redshift z respectively),  we estimated the virial radius of $r_{200}$=1.3 Mpc i.e., $r_{500}\sim1$ Mpc (using X-ray mass of $2.9\times10^{14}\;\rm{M_{\odot}}$) as shown in Figure~\ref{fig:x-ray}. Further, our estimate for the average temperature of the gas, $T_g$, is 3.5 keV (i.e. $4.0\times10^{7}\rm{K}$) as calculated from the relation $T_{X}=2.34\;L_{44}^{1/2}\;h_{50}$ \citep{Bohringer_2000ApJSb} using $L_{X}=1.5\times10^{44}\;\rm{erg\;s^{-1}}$ for uncorrected X-ray luminosity.

\section{Discussion}\label{discus}

\begin{figure}
\includegraphics[width=9cm]{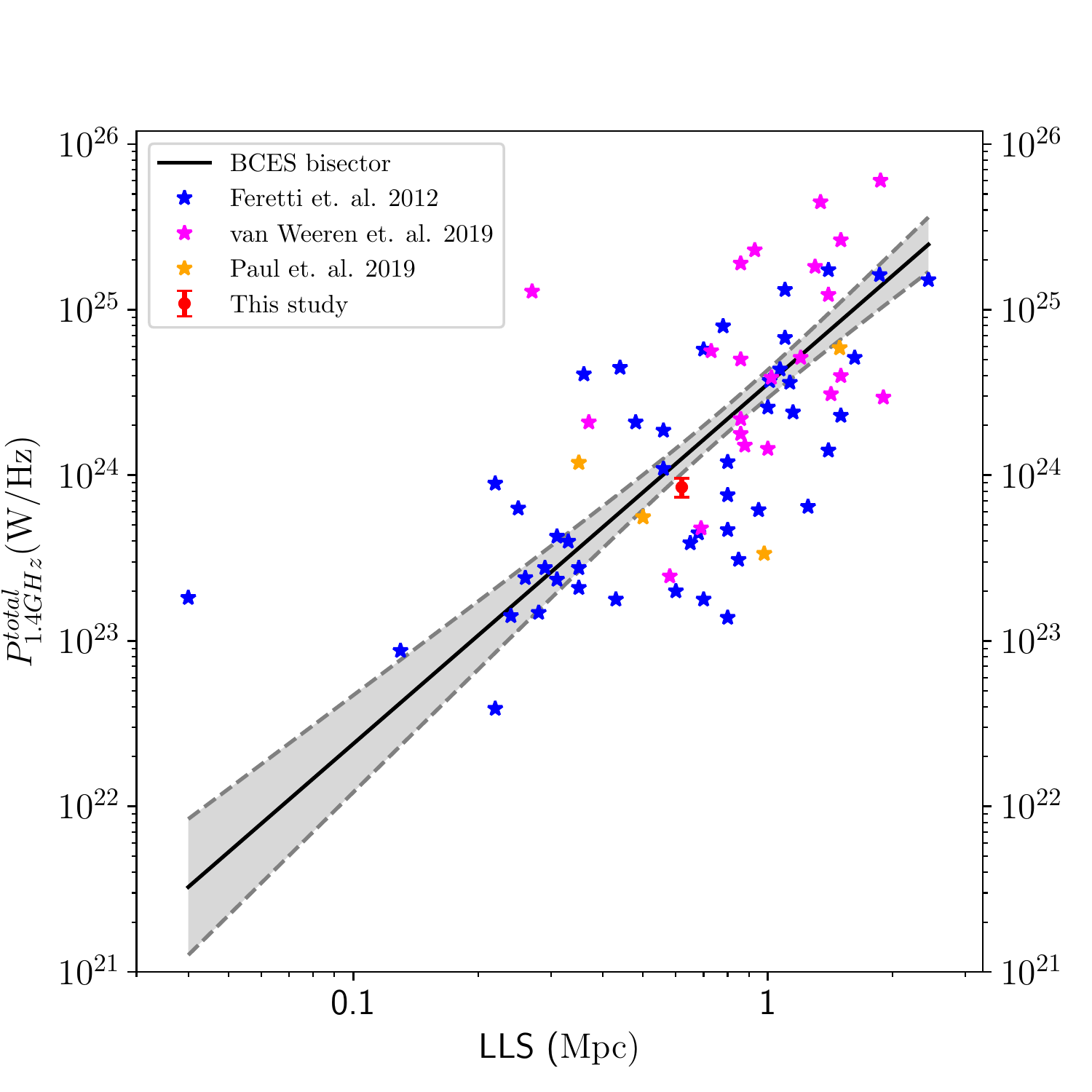}
\caption{BCES-Bisector fit for the radio power (1.4 GHz) vs LSS correlation. Data collected from \citealt{Feretti_2012A&ARv,Paul_2019MNRAS,Weeren_2019SSRva}. Relic found in Abell 1697 is plotted as red point with error bars.}\label{fig:p14-LLS}
\end{figure}

The elongated, peripheral diffuse radio emission found in the field of cluster Abell 1697 has a largest linear size of 830 kpc and width of 300 kpc in LoTSS map. Extreme end of this proposed relic emission is found at about $r_{500}$ i.e., $\sim$1 Mpc away from the X-ray peak (see Fig.~\ref{fig:x-ray}). The emission is purely diffuse in nature as evident from the radio and optical images (see section~\ref{radio-sig}). The average spectral index estimated for this relic is relatively flat ($\alpha^{1.4 GHz}_{144 MHz}=-0.96\pm0.17$), but well within the variations found in relics so far \citep{Weeren_2019SSRv}. The average equipartition magnetic field estimated (see the method in \citealt{Jamrozy_2004A&A}) as $0.6~\mu$G. The spectrum of the relic is flatter at the front i.e. at injection ($\alpha_{inj}=-0.70\pm.14$) and gradually steepens to ($\alpha_{edge}=-1.19\pm0.23$) at the inner edge, indicating a Compton synchrotron emission process with continuous injection \citep{Bonafede_2012MNRAS}, usually a characteristic of peripheral relics. Further, we found that the proposed relic perfectly follow the correlation of relic radio power (at 1.4 GHz) and its size (LLS) with a correlation slope of $2.17\pm0.25$ (see the plot in Fig~\ref{fig:p14-LLS}), first discussed in \citet{Feretti_2012A&ARv} and recently updated in \citet{Paul_2019MNRAS}.

From the ROSAT survey X-ray map (i.e., Fig.\ref{fig:x-ray}), we detect features indicative of ongoing dynamical activities in the cluster. The  contours from the X-ray photon count map coincide well with the radio relic morphology (see Fig.\ref{fig:x-ray} and section~\ref{x-ray}). Also strikingly, a dip in X-ray photon count observed in the profile map at a distance of $850\pm70$ kpc from the X-ray peak. This potential shock is in the sector that encloses the radio relic (see Fig.~\ref{fig:x-ray1}~(b)), and reasonably matches with the distance between the X-ray peak to the brightest radio emission (i.e. 815 kpc). The relic front (i.e. 970 kpc), as well as its extension ($\sim710-970$ kpc), indicates a connection between the peripheral radio emission and the X-ray shock. An inadequate photon count restricted us from making a temperature profile to confirm this connection.

\subsection{Turbulent re-acceleration by wake turbulence?}

We performed a cosmological simulation for a merging cluster using {\sc ENZO}, an N-Body plus hydrodynamic, Adaptive Mesh Refinement (AMR) code \citep{Bryan_2014ApJS}. A realization of a sky patch with a linear scale of 128 Mpc $(1+z)^{-1} h^{-1}$ \;[i.e.  having a co-moving volume of $(128 \; \rm{Mpc\;h^{-1}})^3$] is produced in a flat $\Lambda$CDM background cosmology. The radio emission is computed from the simulated merging cluster implementing synchrotron emission mechanism invoking both the DSA and Turbulent Re-connection Acceleration (TRA) as the electron injection models to understand the large radio emission found in Abell 1697. For more details of this simulation studies, as well as the time series of cluster merger events and its thermal and non-thermal energy evolution, please see Appendix~\ref{radio-model}. 

In Figure~\ref{fig:sim-rad-x}(a), we see a post merger thermal shock producing the wake turbulence behind it (see Fig.~\ref{fig:sim-rad-x}(b)). The simulated mock radio as well as X-ray map (using {\sc{cloudy}} code, \citet{Ferland_2017RMxAA}), shown in Figure~\ref{fig:sim-rad-x}(c), indeed show a possibility of finding a peripheral relic due to DSA along with a trailing radio emission extending till the cluster centre due to TRA, thus tentatively explaining the observed radio structures reported in this study.

\begin{figure*}
\includegraphics[width=6.8cm]
{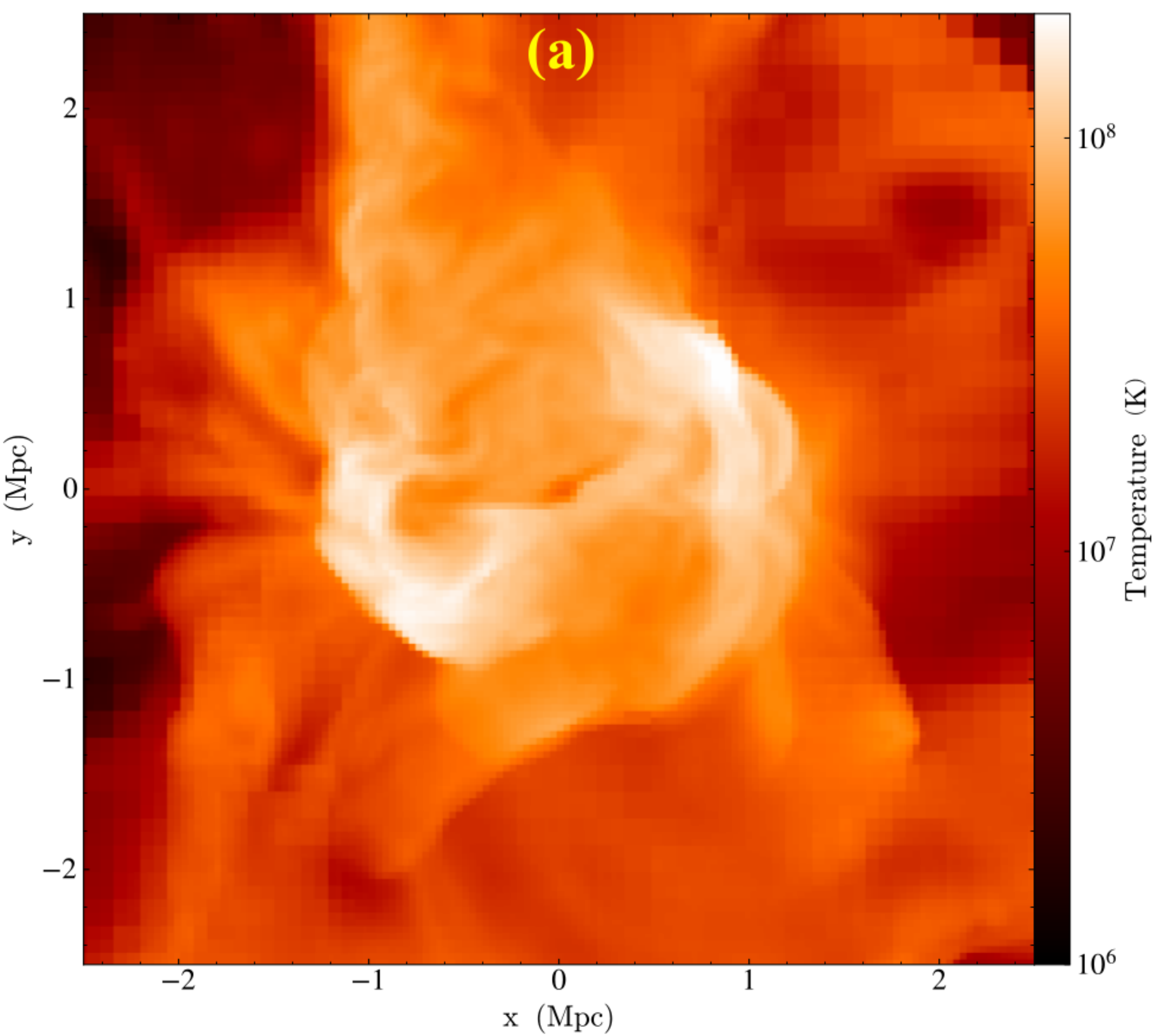}\hspace{-0.5cm}
\includegraphics[width=6.8cm]
{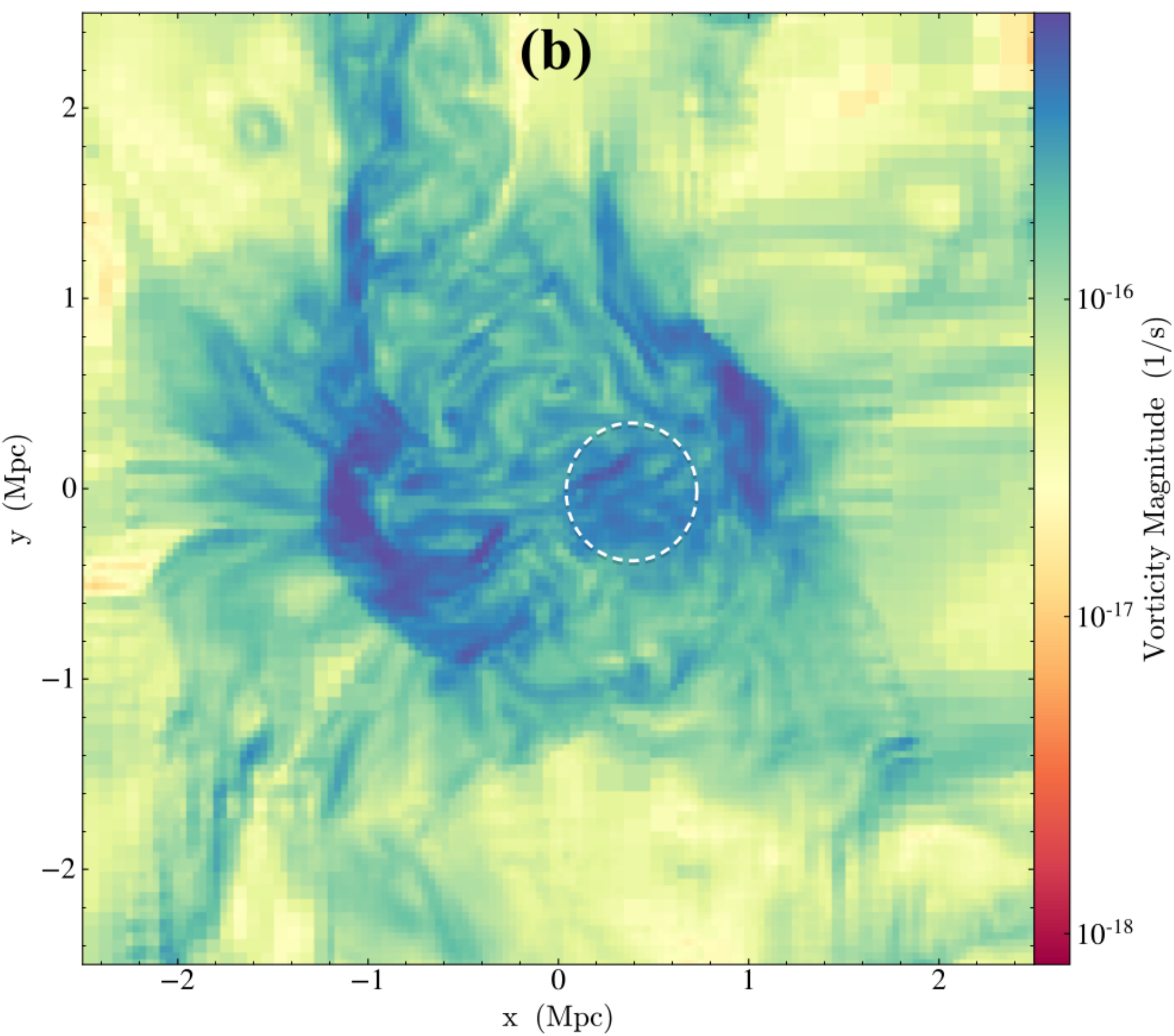}\hspace{-0.5cm}
\includegraphics[width=6.2cm]{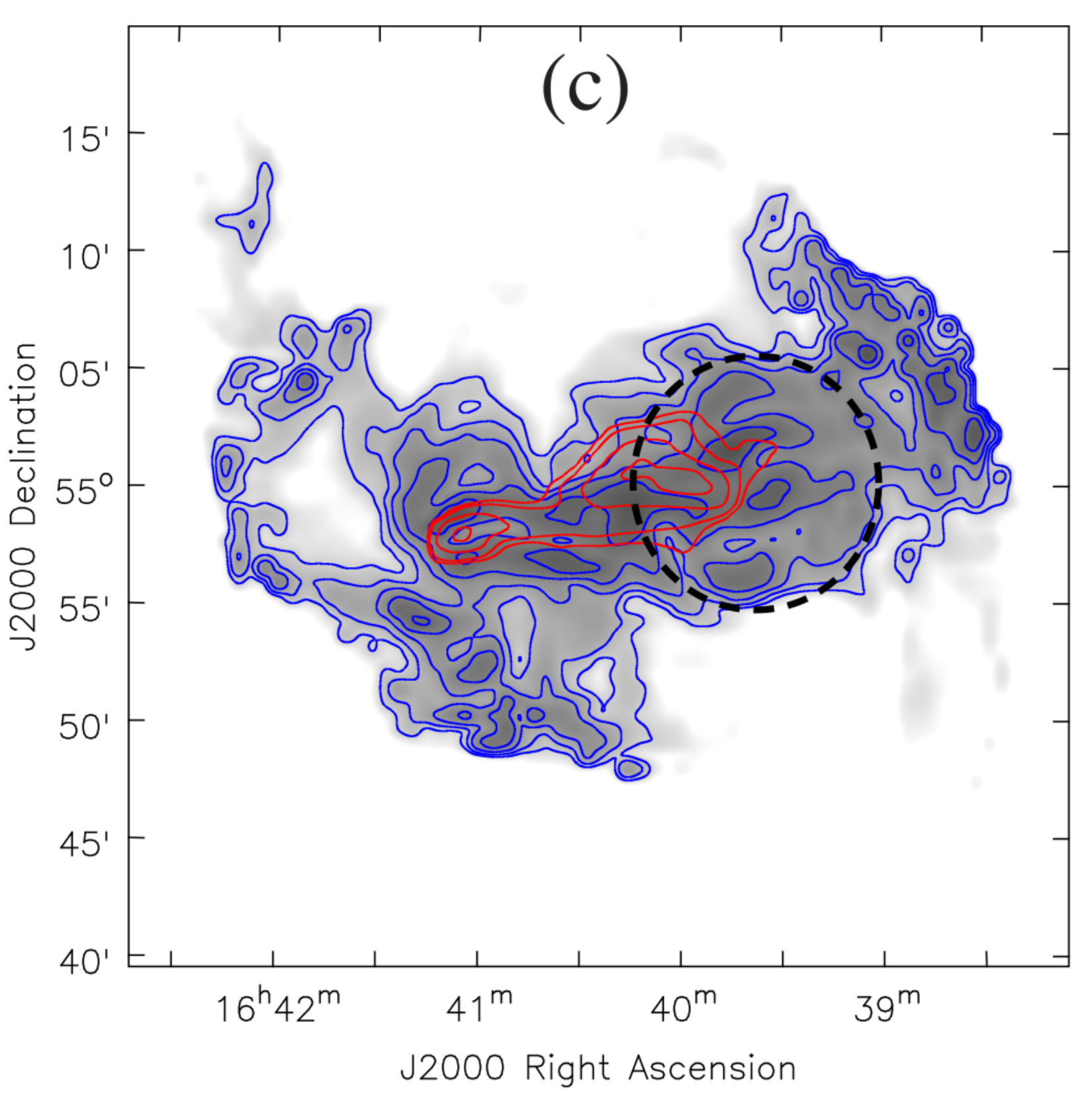}\hspace{-0.8cm}\\
\caption{{\bf Panel~a:} and {\bf Panel~b:} shows respectively the post-merger temperature shock and the vorticity magnitude ($\bar{\omega}=\nabla\times\bar{\rm{v}}$ (representing turbulence) of a simulated cluster. {\bf Panel~c:} Modelled mock radio map considering radio emission from both DSA and TRA electrons (see Appendix~\ref{radio-model}) plotted as gray (zoomed on the cluster). The computed X-ray luminosity is overlaid in red contours. Co-ordinates shown are arbitrary.}\label{fig:sim-rad-x}
\end{figure*}

\subsection{A radio lobe angle?}

Giant radio galaxies usually have two symmetric radio lobes. Also, the brightest emission observed inside the proposed relic does not resemble the hot-spots of a radio galaxy in morphology, magnetic field strength (much lower at $\sim 1.4~\mu$G), as well the unusually steep spectrum ($\alpha<-1$). The trailing diffuse emission shows an extremely steep average spectrum (upper limit) of $\alpha_{trail} \leq -1.84$ with an average magnetic field and age of the electrons computed as $0.8~\mu$G and $190$~Myr respectively (using the method of \citealt{Jamrozy_2004A&A}), hardly fits the characteristics of an active giant radio galaxy. Further, the brightest source (i.e. `S1' in Fig.~\ref{fig:radio-lofar}) does not seem to be just a point source as indicated from its steep spectrum ($\alpha<-1$). Indeed, the VLA FIRST map resolves it to two prominent radio sources (Fig.~\ref{fig:optical-overlay}, inset). One is a bright radio double source with a central optical counterpart having VLA FIRST (1.4 GHz) flux density of $40.4\pm4.1$ mJy. The two other optical sources observed near to each of the radio lobes do not match the radio peak positions. One among them in the NE direction is not even marked as a galaxy in NED. Therefore, the central radio source is a confirmed double source, surely not the core of a giant radio galaxy. The second radio source S3, is a very weak one with VLA FIRST 1.4 GHz radio power of $P_{1.4\;\rm{GHz}}=5.24\pm1.26\times10^{23}\;\rm{W\;Hz^{-1}}$, unlikely to host a giant radio source of size more than a Mpc \citep{Lara_2004A&A}.

\subsection{Is the trailing diffuse emission a radio phoenix?}

The other feasible situation is that the structure is a revived radio lobe. The spectral index upper limit of the trailing diffuse radio emission ($\alpha_{144 MHz}^{1.4 GHz} (trail) < -1.84$ \& $\alpha_{144 MHz}^{330 MHz} (trail) < -2.02$) indicates it likely falls in the category of ultra-steep spectrum radio sources. A cluster merger shock may have passed through the ghost electron clouds from the past AGN activity and revived the fossil electrons by adiabatic shock compression, as in the scenario proposed by \citealt{Ensslin_2001A&A}, called the radio phoenixes. The proposed phoenix has a large size of about $800\times550$~kpc with extremely faint radio surface brightness of $0.3~\mu$Jy arcsec$^{-2}$ at 144 MHz LoTSS map. The estimated magnetic field strength is low, with an average of $\sim0.8~\mu$G and the average electron age of $190$~Myrs. The distance of the phoenix centre to the shock front is about 500~kpc. With a shock speed of $\sim2020~\rm{km~s^{-1}}$, as computed from the average shock Mach number of $\mathcal{M}\sim2.3$ and the temperature of $\sim3.5$~keV (from the relation $V_{sh}=\mathcal{M}[1480(T_g/10^8 K)^{1/2}]\;\rm{km\;s^{-1}}$ \citep{Sarazin1988book} where $T_g$ is the temperature of the downstream ICM gas), the shock passage time is about 250~Myrs, thus indicating a strong case of revival of fossil electrons to generate the trailing diffuse radio emission found in cluster Abell 1697 at 144 MHz LoTSS maps.

\section{Conclusions}\label{conc}

We report the discovery of a relic-like diffuse radio emission from the outskirts of the cluster Abell 1697 as well as an extremely steep spectrum diffuse radio emission trailing behind the said relic. LoTSS, NVSS and WENSS radio maps, along with the X-ray map from the ROSAT All-Sky Survey, provide reasonable evidence that the structure is a peripheral relic. The trailing diffuse radio emission has very low surface brightness and an extremely steep spectrum, indicative of a radio phoenix.

We discussed various possibilities to understand the mechanism responsible for the observed radio structures invoking turbulent re-acceleration as well as the process of revival of fossil electron clouds from an old AGN activity. However, no firm conclusions on its origin can be drawn due to the absence of deep observations, mainly at high radio frequencies and in X-rays.

\begin{acknowledgements}:
This research was funded by DST INSPIRE Faculty Scheme awarded to Dr. Surajit Paul (code: IF-12/PH-44). S. Salunkhe wants to thank ``Bhartratan JRD Tata Gunwant Sanshodhak Shishyavruti Yojna" for providing doctoral fellowship. PG acknowledges Council of Scientific \& Industrial Research (CSIR) for providing senior research fellowship (CSIR-SRF) for supporting his PhD work. We are thankful to Dr. Neelam Panwar of ARIES, Naitital for her valuable suggestion regarding optical and UV image interpretation. SP and PG would like to thank The Inter-University Centre for Astronomy and Astrophysics (IUCAA) for providing the HPC facility. Computations described in this work were performed using the Enzo code developed by the Laboratory for Computational Astrophysics at the University of California in San Diego (\url{http://lca.ucsd.edu}) and data analysis of the simulation is done with the yt-tools (\url{http://yt-project.org}).
\end{acknowledgements}

\begin{appendix}

\section{Modelling of cluster radio emission through cosmological simulations}

A cosmological simulation for a merging cluster was performed using {\sc ENZO}, an N-Body plus hydrodynamic, Adaptive Mesh Refinement (AMR) code \citep{Bryan_2014ApJS}. A flat $\Lambda$CDM background cosmology with $\Omega_\Lambda$ = 0.7257, $\Omega_m$ = 0.2743, $\Omega_b$ = 0.0458,  h = 0.702 and primordial power spectrum normalization $\sigma_8$ = 0.816 derived from WMAP \citep{Komatsu_2009ApJS} were used. The simulation was initialized at redshift $z = 60$ using the \citet{Eisenstein_1999ApJ} transfer function, and evolved up to redshift $z = 0$. An ideal equation of state was used for the gas, with $\gamma = 5/3$. Since the emergence and propagation of shocks are the most important events in this study, we thus captured the shocks very efficiently and resolved the grids adaptively wherever the shock is generated and passes by using the method described in \citet{Vazza_2009MNRAS}. In order to capture the correct energy distribution of the ICM, radiative cooling \citep{Sarazin_1987ApJ} 
and a star formation feedback scheme was applied \citep{Cen_1992ApJS}. Radio emission modelling was done as the post-processing numerical analysis by using the yt-tools \citep{Turk_2011ApJS}. A detailed description of the simulation can be found in \citet{Paul_2018arXiv}. 

Cosmological simulations were performed to create a sky realization with a linear scale of 128 Mpc $(1+z)^{-1} h^{-1}$ \;[i.e.  having a co-moving volume of $(128 \; \rm{Mpc\;h^{-1}})^3$] with 0.3 million particles and 64$^3$ grids at the root grid level. We further introduced 2 child grids and inside the innermost child grid of $(32\; \rm{Mpc\;h^{-1}})^3$ co-moving volume, we implemented 4 levels of AMR based on both over-densities and the shock strength. With this set-up, we achieved about 30 kpc h$^{-1}$ co-moving spatial resolution and mass resolution of about 10$^{8} M_{\odot}$ at the highest resolved grids. For further details of the simulations and resolution studies of our numerical schemes, readers are suggested to go through \citet{Paul_2017MNRAS,Paul_2018arXiv}.

\subsection{Radio emission modelling}\label{radio-model}

\begin{figure*}
\includegraphics[width=6.5cm]{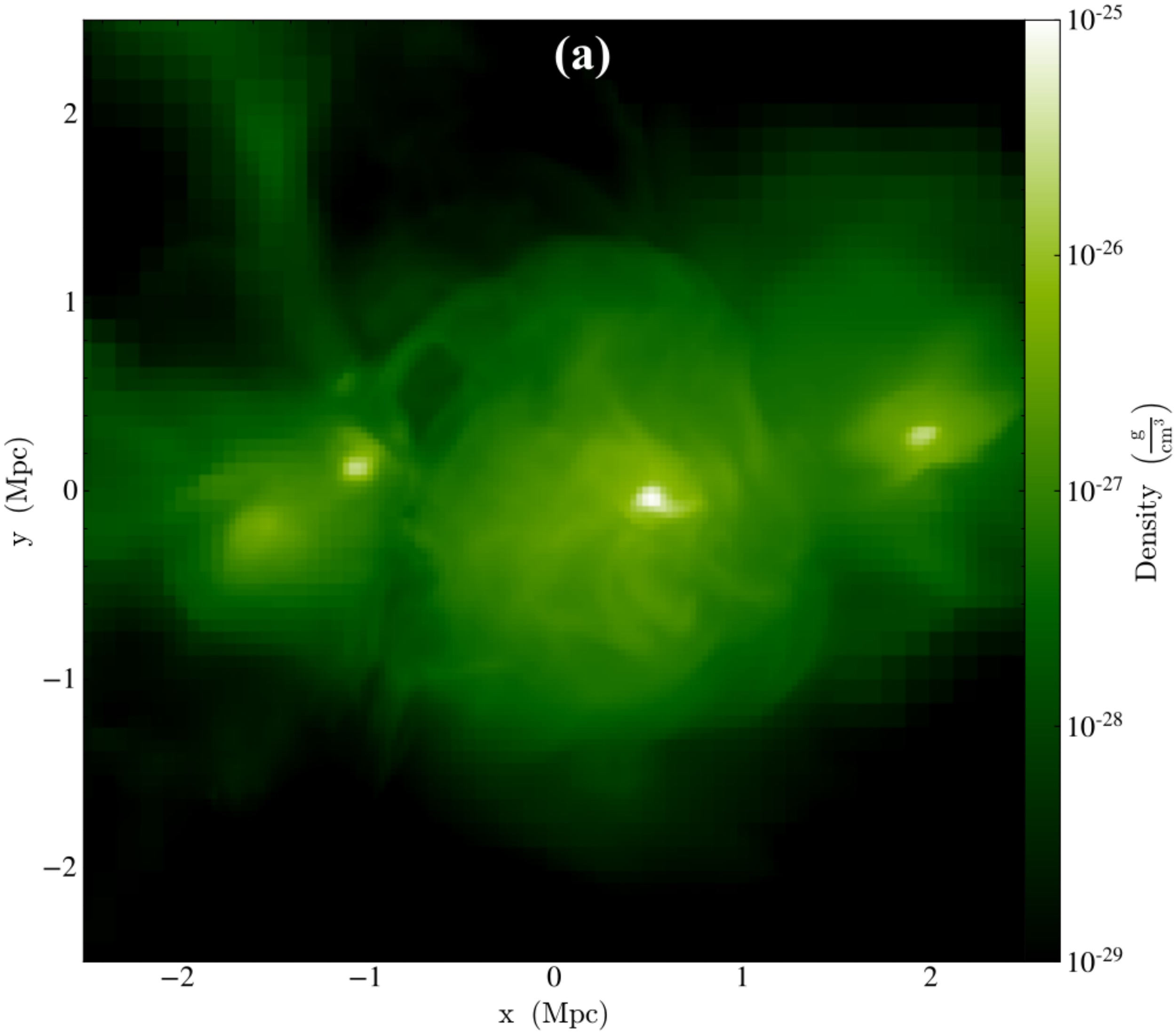}\hspace{-0.9cm}
\includegraphics[width=6.5cm]
{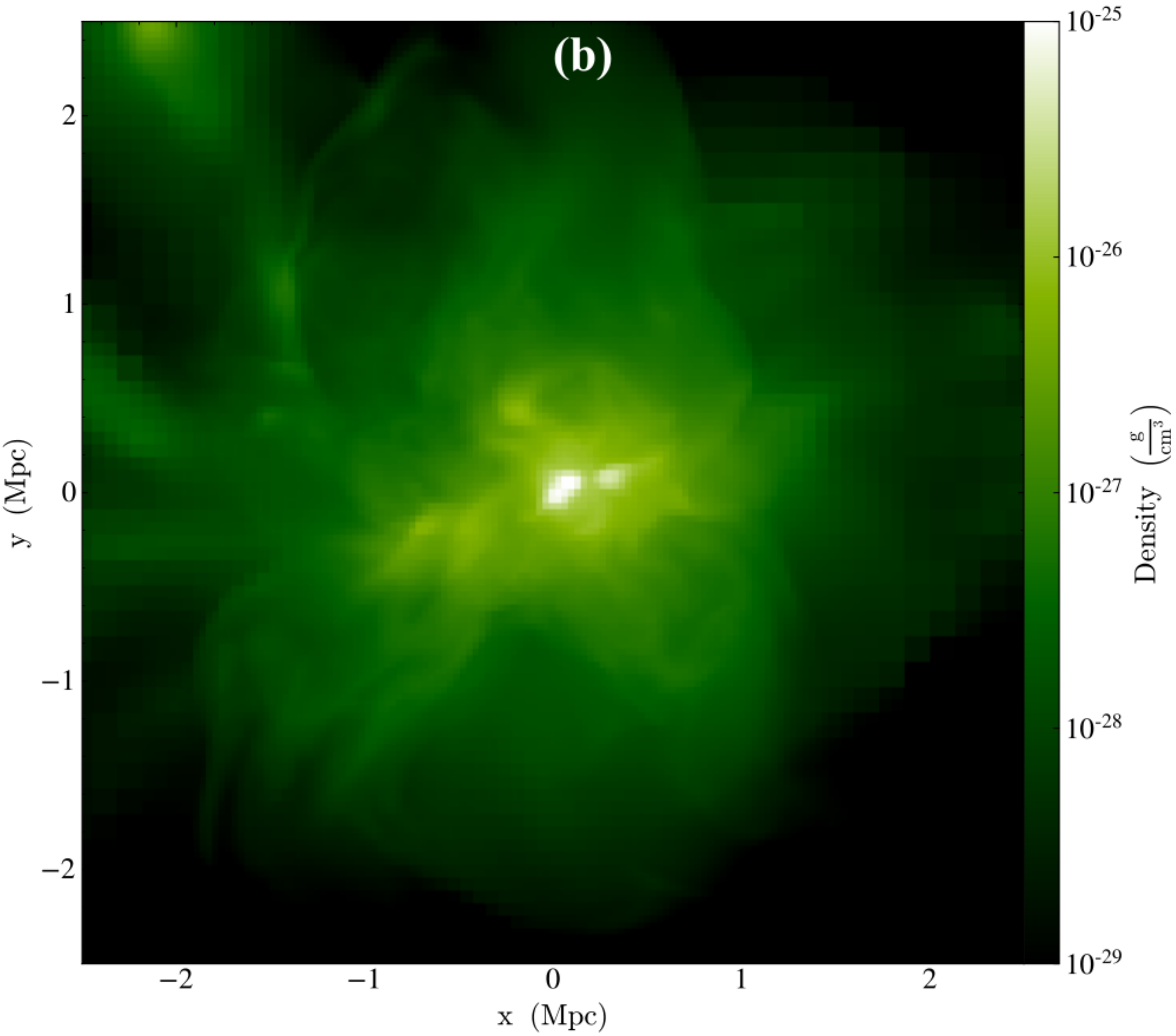}\hspace{-0.9cm}
\includegraphics[width=6.5cm]
{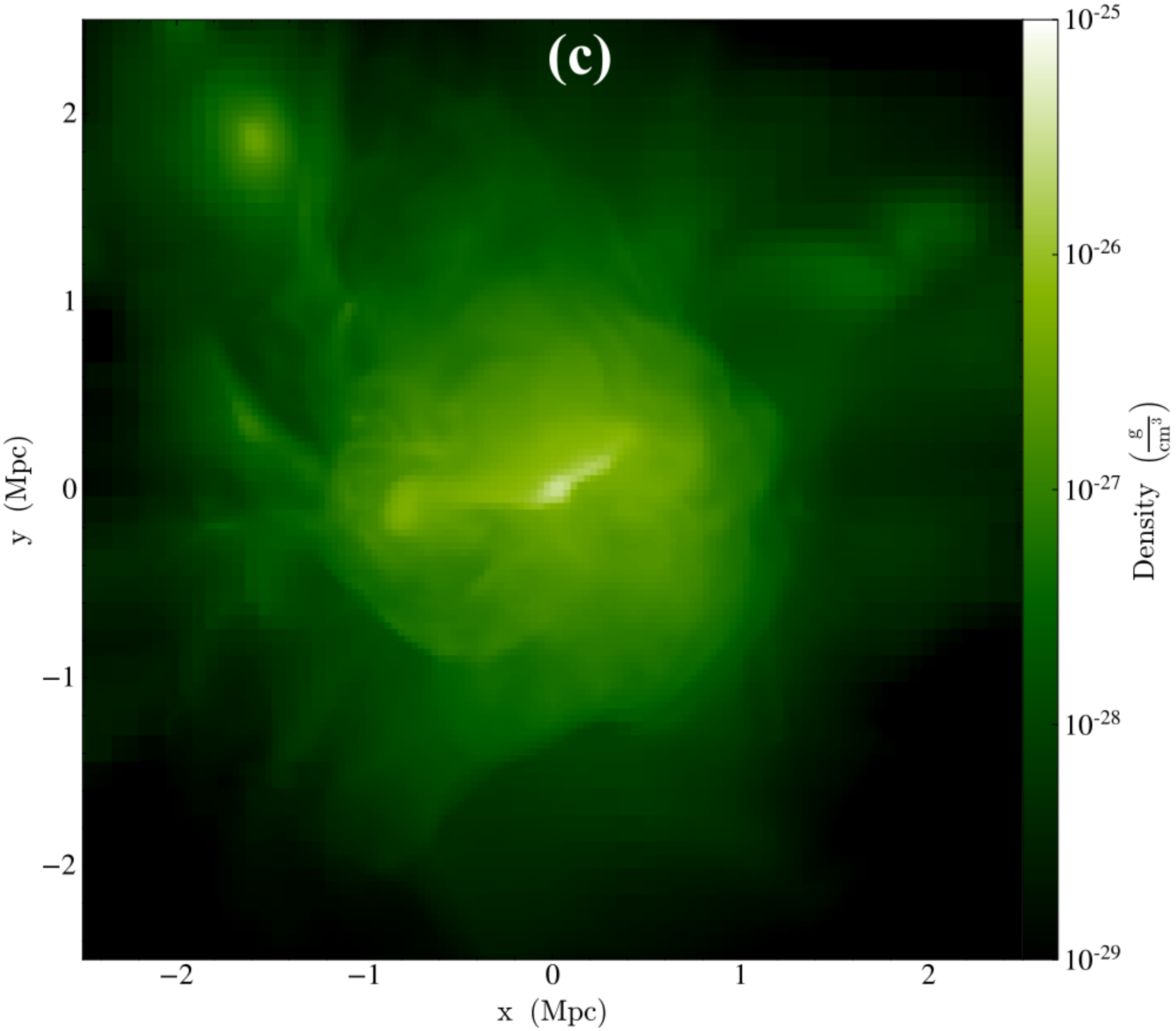}\hspace{-0.8cm}\\
\includegraphics[width=6.5cm]{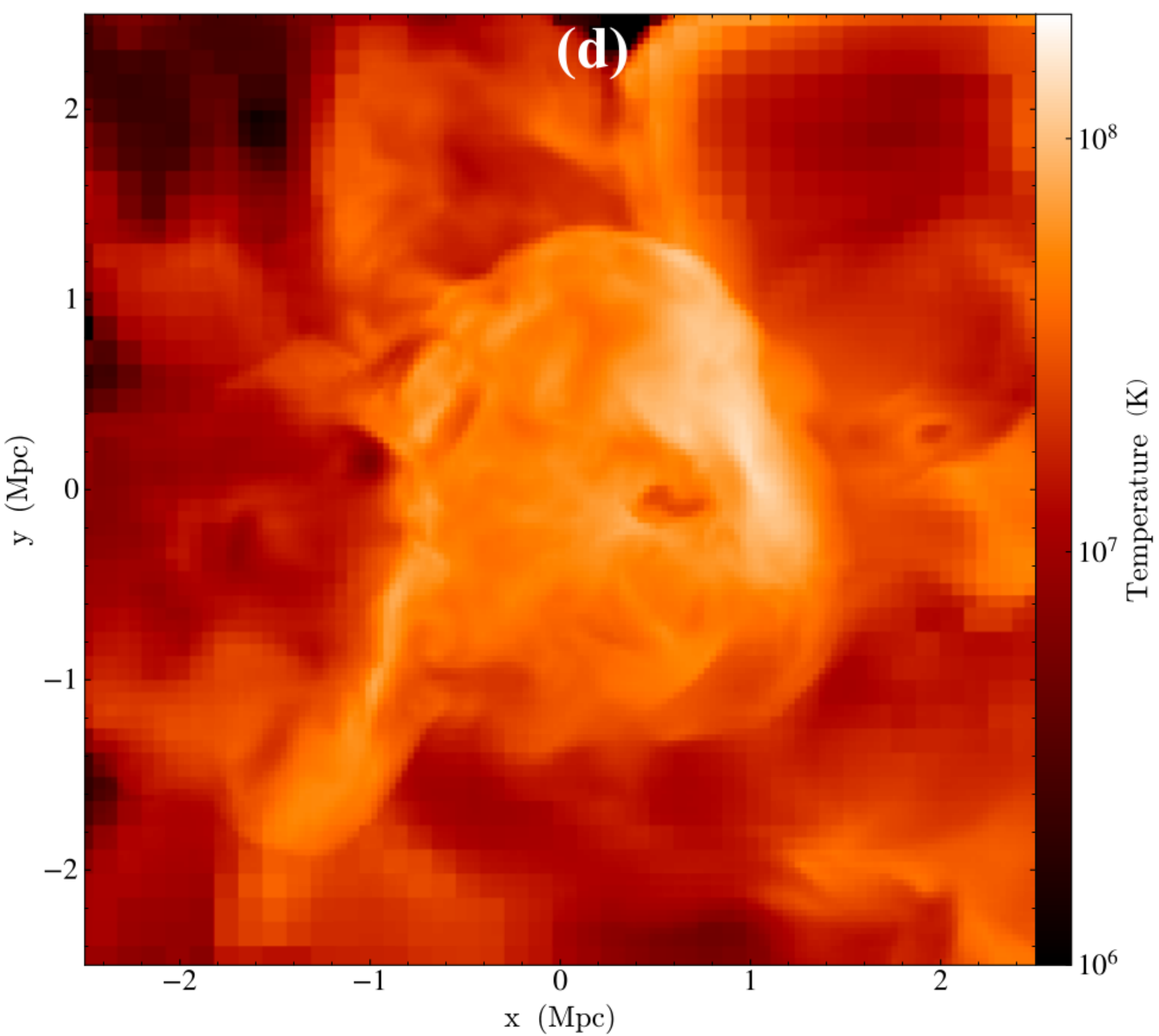}\hspace{-0.9cm}
\includegraphics[width=6.5cm]
{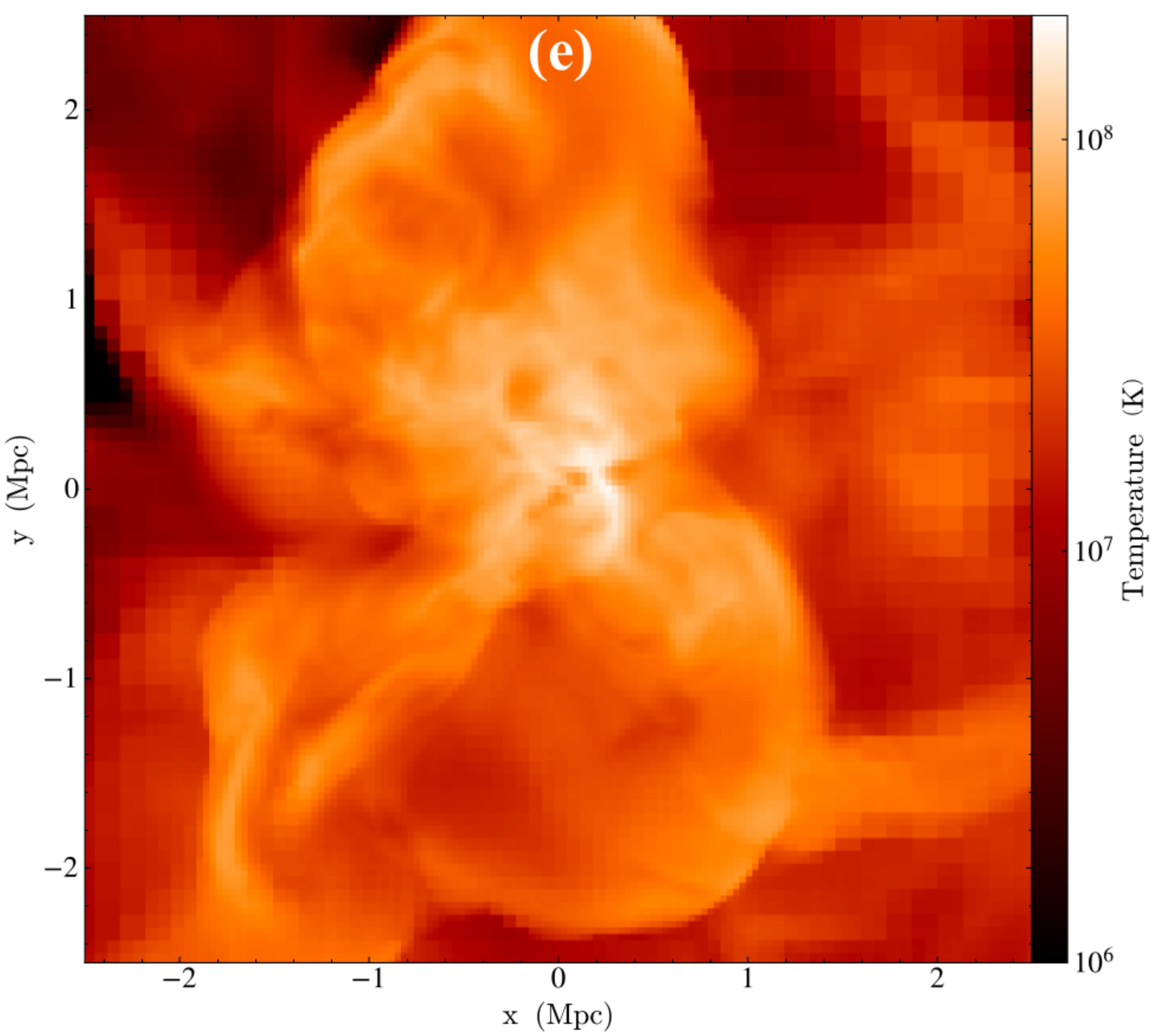}\hspace{-0.9cm}
\includegraphics[width=6.5cm]
{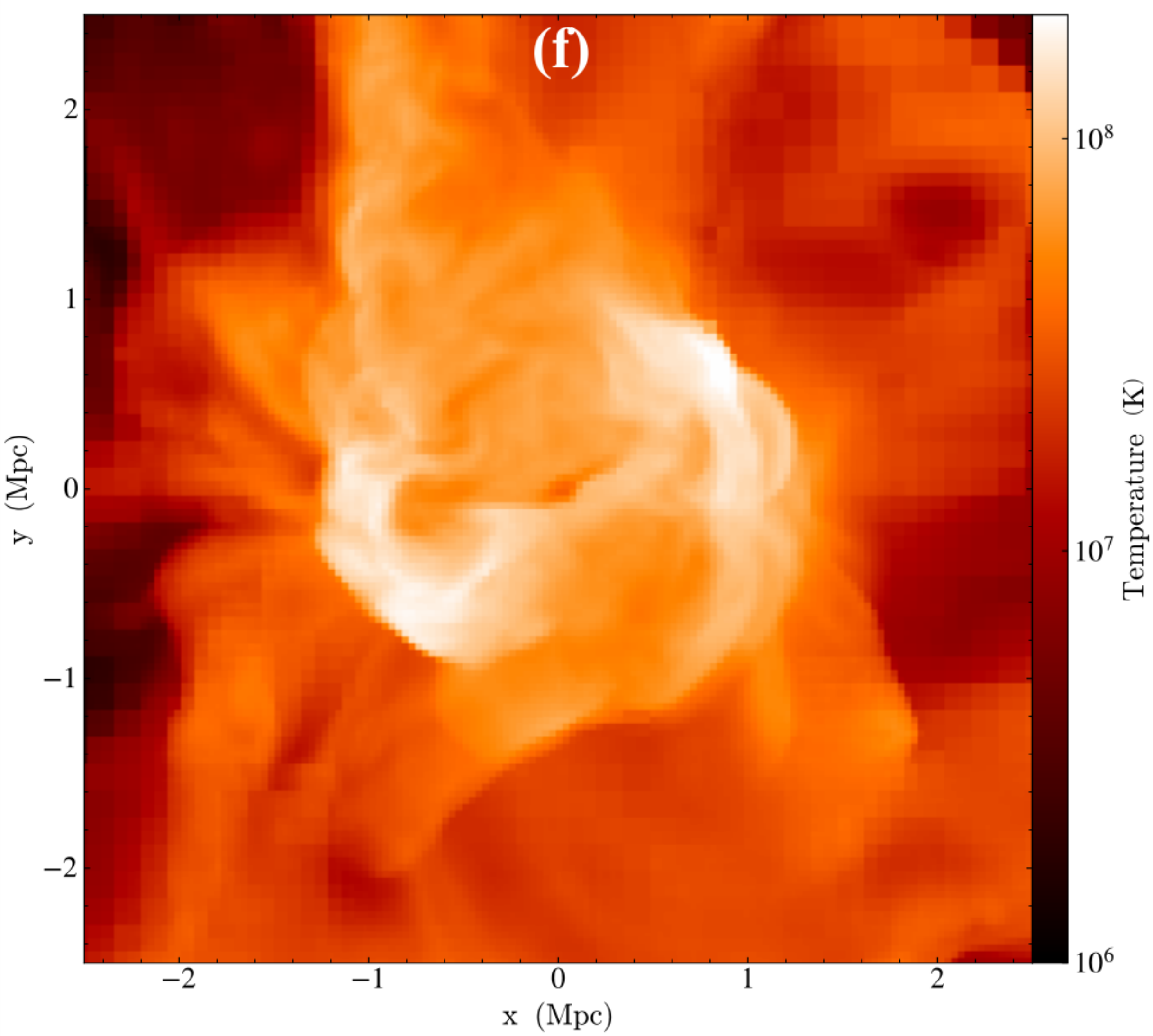}\hspace{-0.8cm}\\
\includegraphics[width=6.5cm]{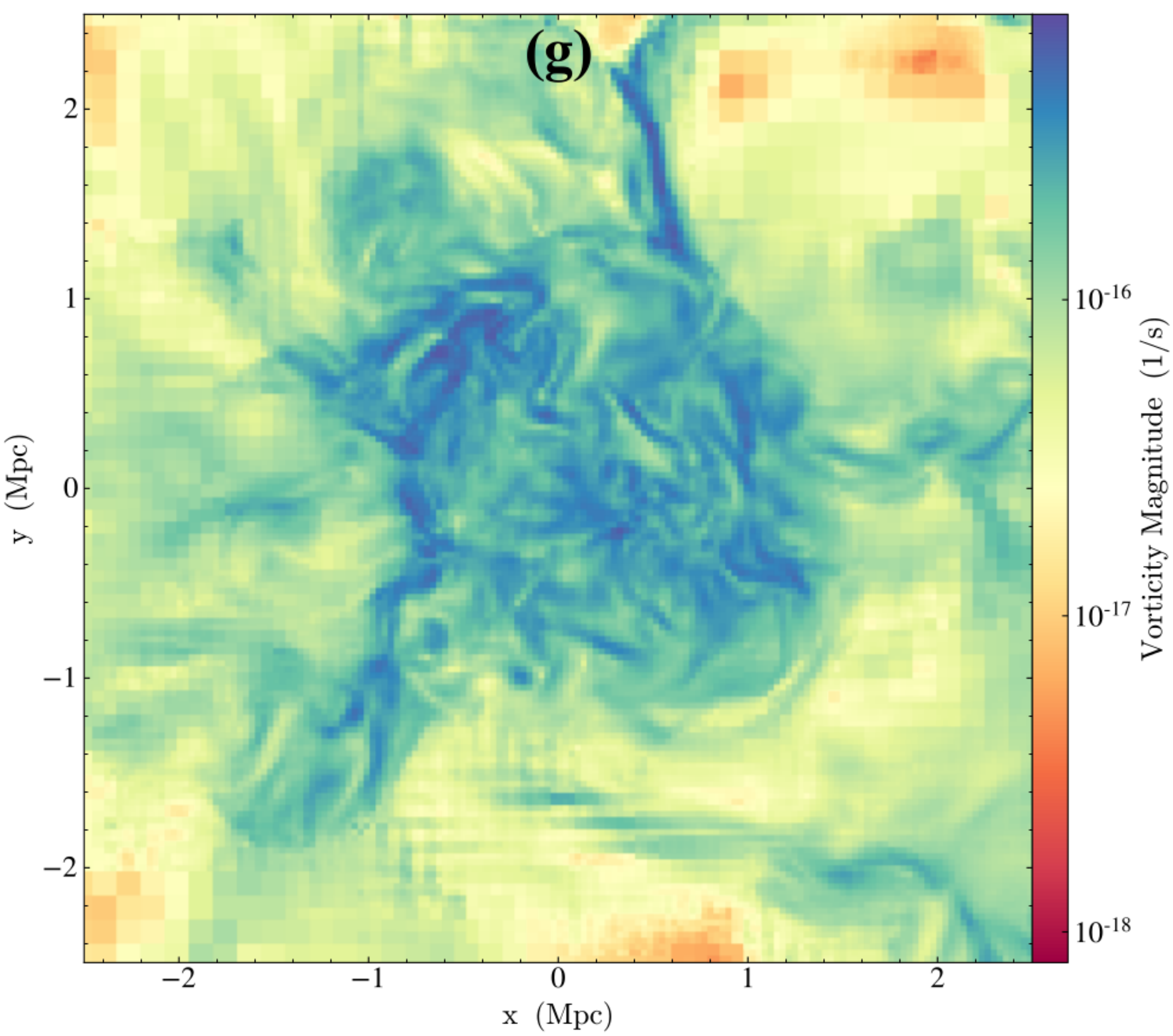}\hspace{-0.9cm}
\includegraphics[width=6.5cm]
{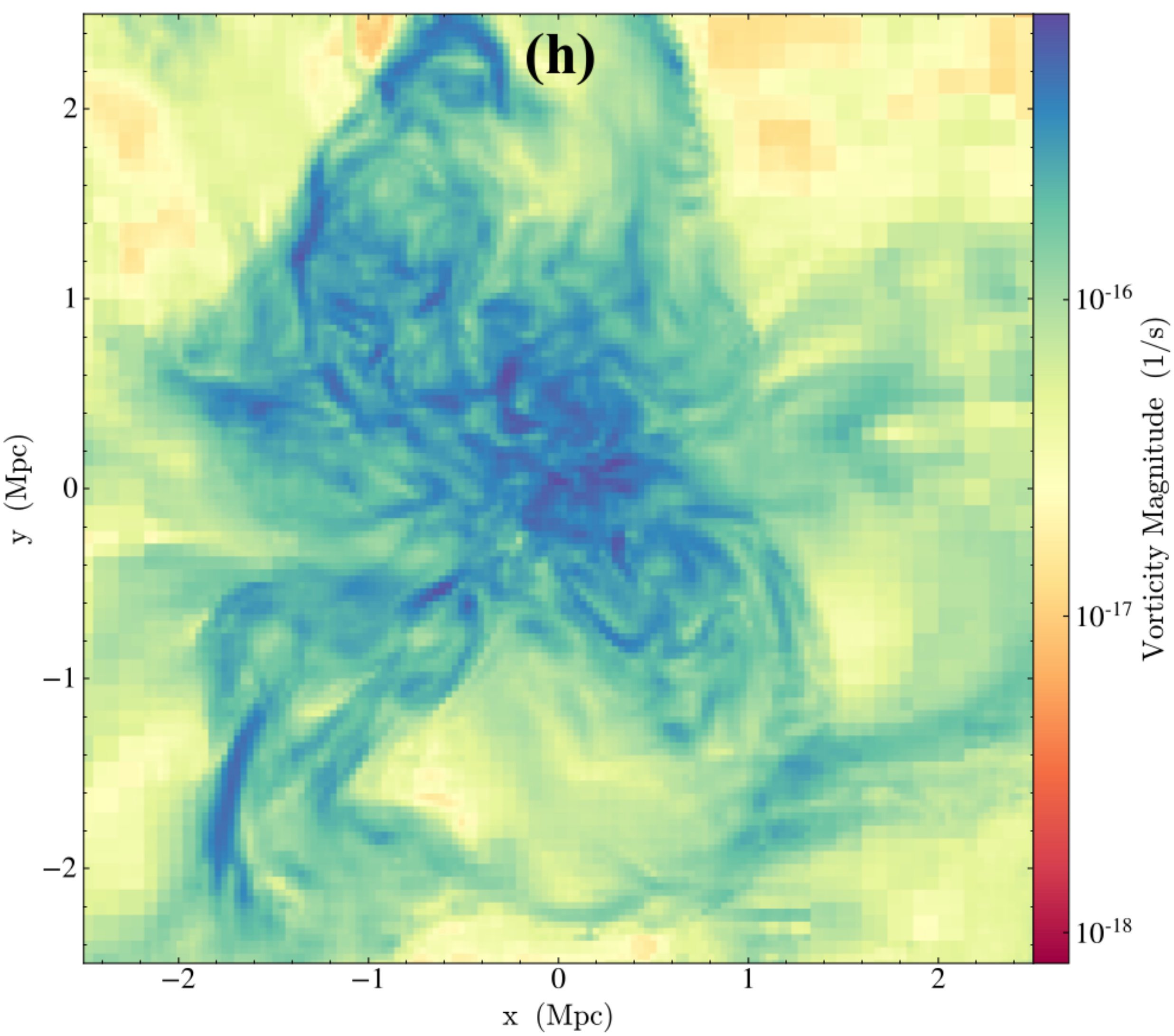}\hspace{-0.9cm}
\includegraphics[width=6.5cm]
{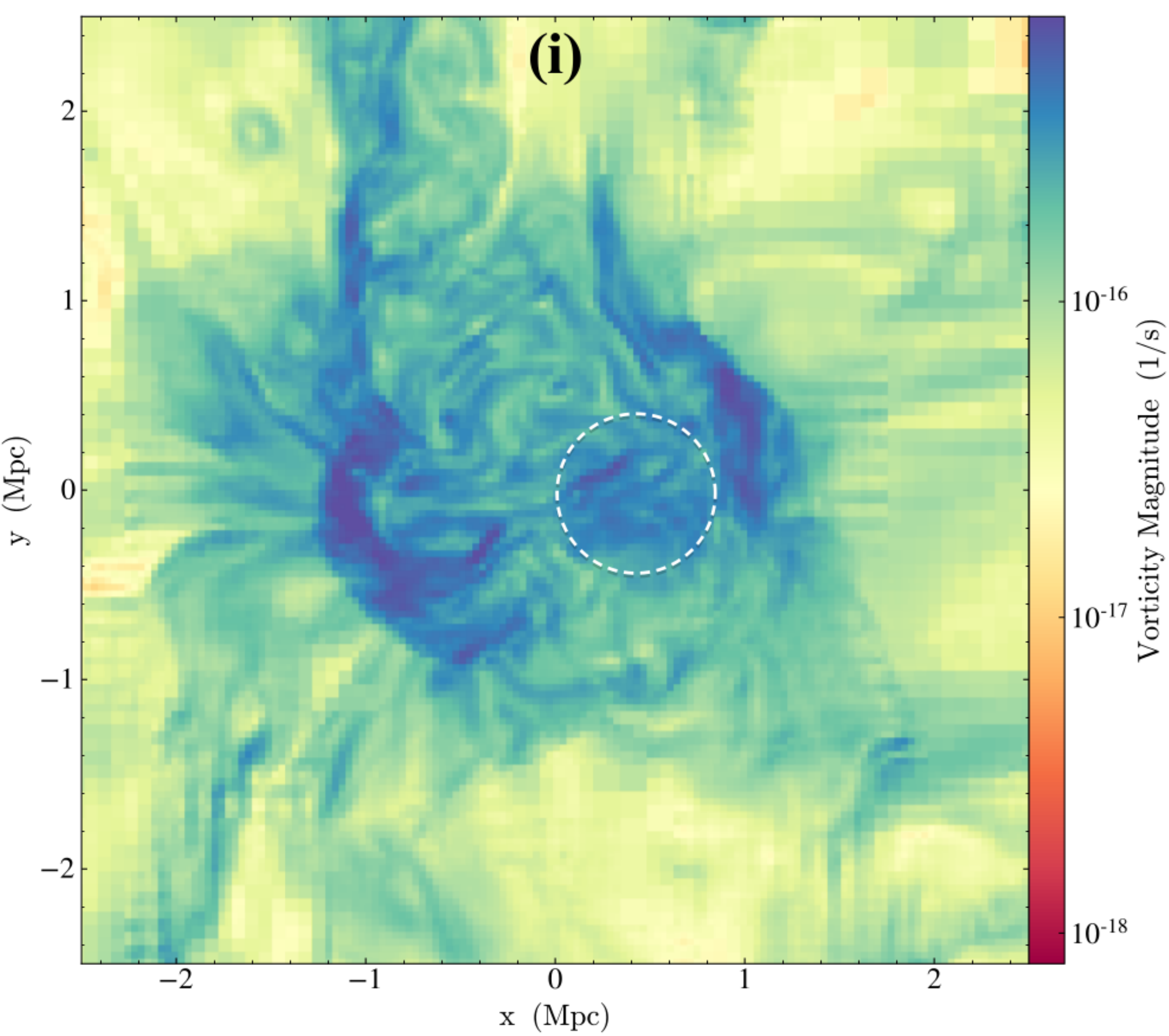}\hspace{-0.8cm}\\
\includegraphics[width=6.5cm]
{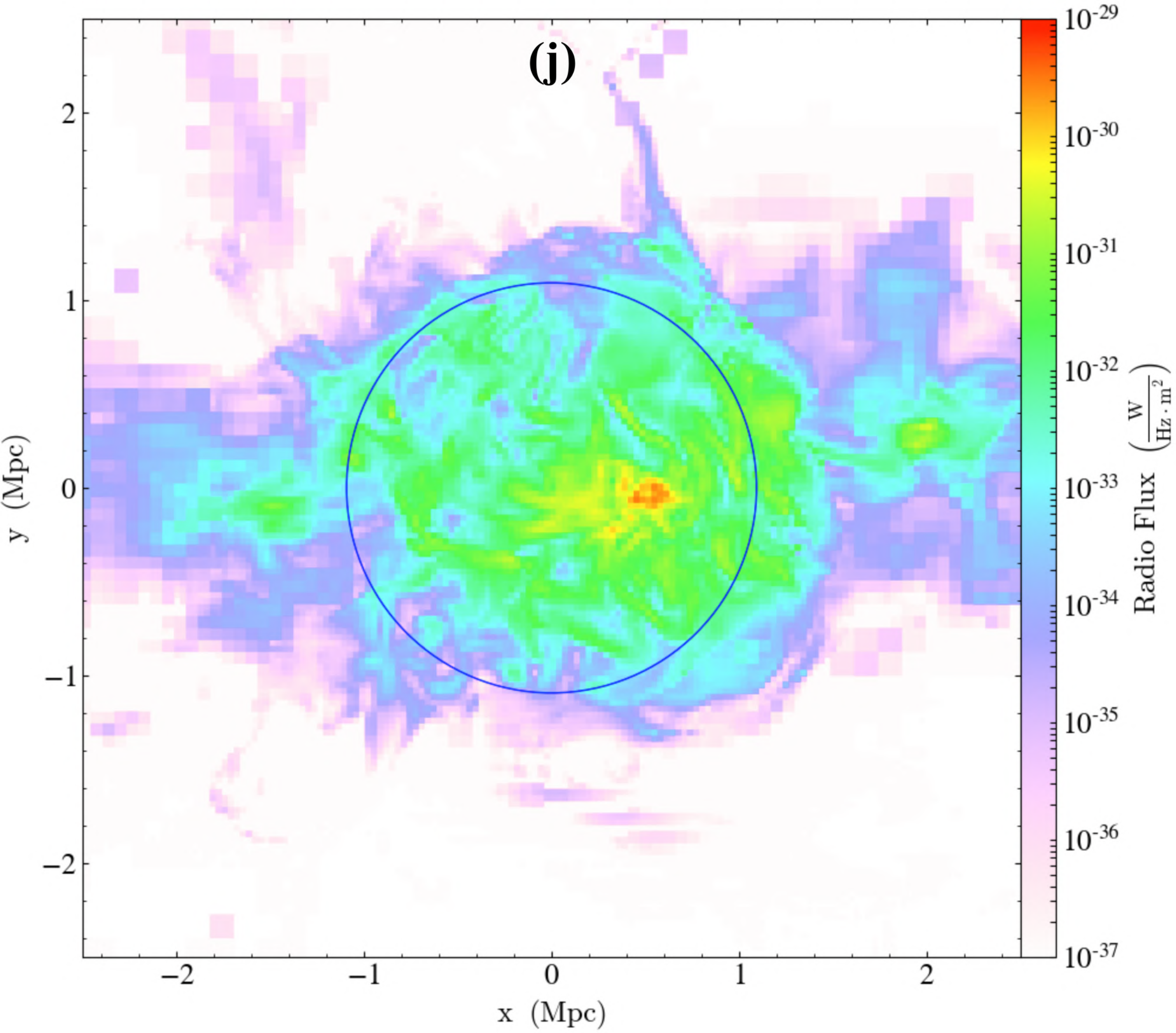}\hspace{-0.9cm}
\includegraphics[width=6.5cm]
{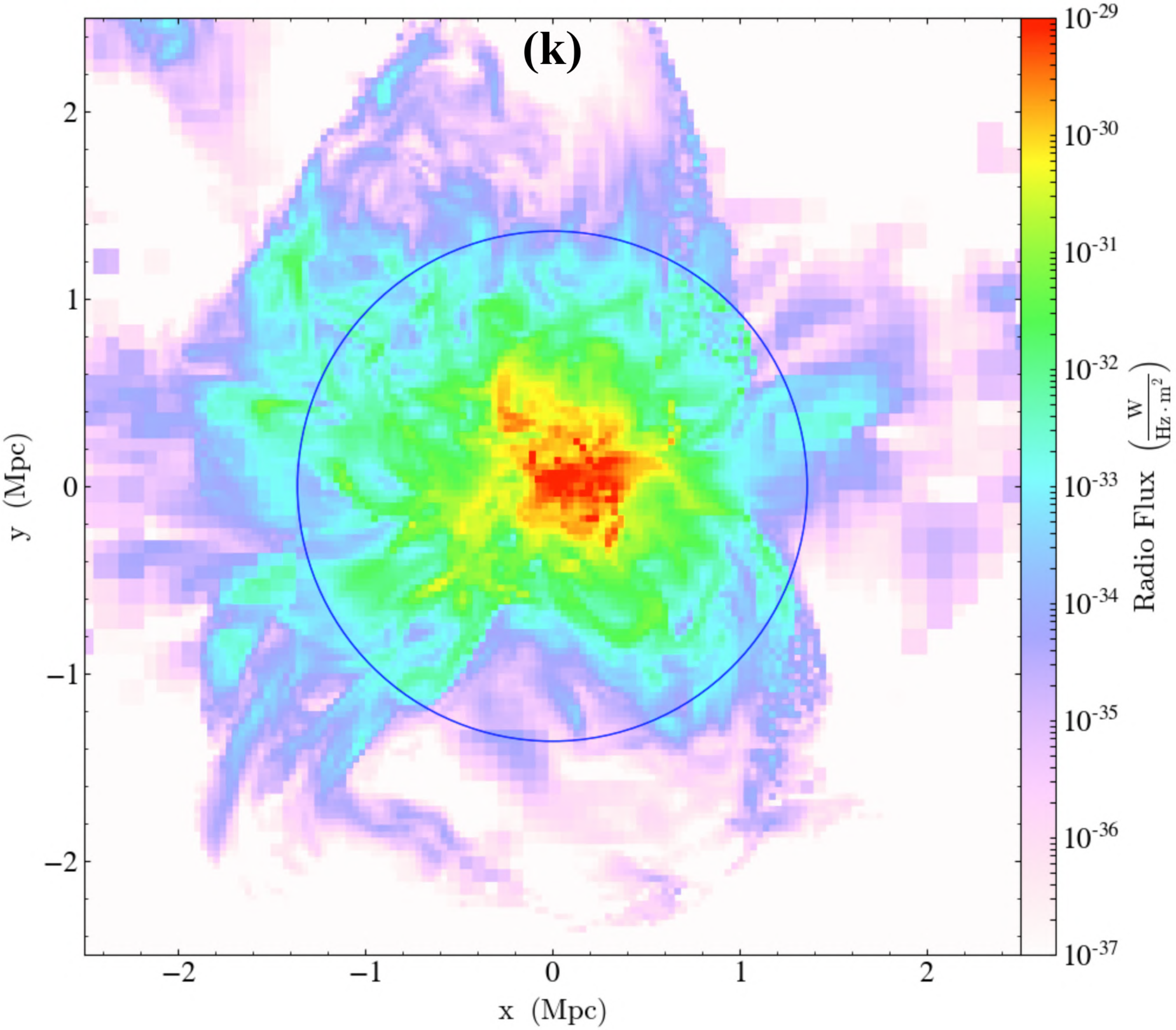}\hspace{-0.9cm}
\includegraphics[width=6.5cm]
{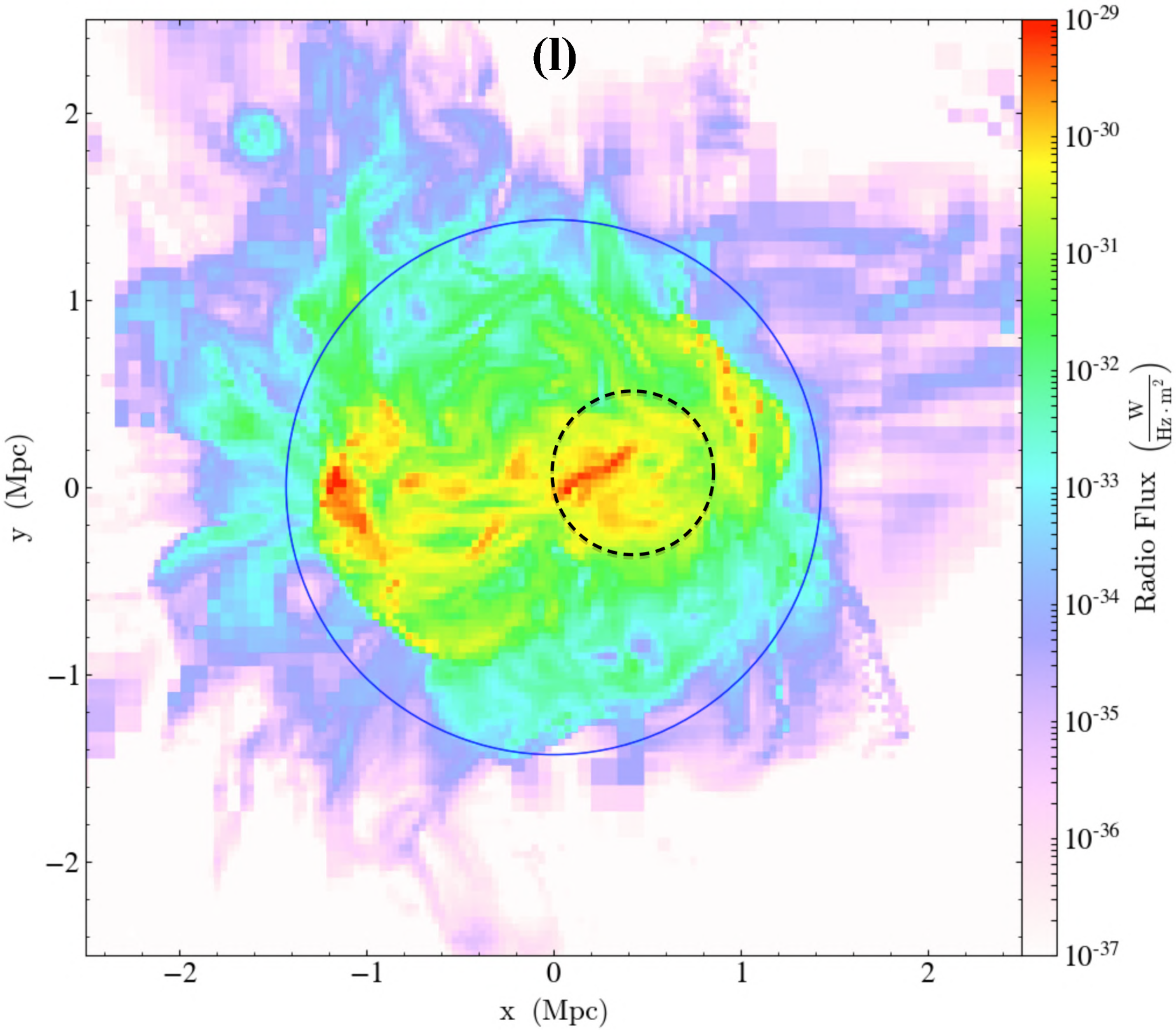}\hspace{-0.8cm}\\
\caption{Each panel is a slice plot of a $5\times5$ Mpc$^2$ area focused on a time series of cluster merger event. Density (first row), Temperature (second row), Vorticity magnitude (third row) and Radio emissions maps of simulated cluster at a pre-merger (first column), merger (second column) and post merger (third column) state. The while and black dashed circle in Panel~(i)\&(l) respectively are wake turbulence and corresponding radio emission from TRA electrons.}\label{fig:radio-emission}
\end{figure*}

Radio emission takes place in a magnetised medium if relativistic charged particles (usually the electrons) are available. The ICM is known to host magnetic field of $\mu$G order \citep{Govoni_2004IJMPD} that can be achieved through turbulent dynamo model applied on primordial seed magnetic field \citep{Subramanian_2006MNRAS}. But, the model is highly dependent on the degree of turbulence in the ICM. A saturation of magnetisation can be obtained once the turbulence fully develops in the medium, usually Kolmogorov type i.e. $E(k) \propto k^{-5/3}$ (where $k$ is wave number, and $E(k)$ is the energy per unit volume in $k$-space). Therefore, the equipartition is considered between magnetic energy density $\frac{B^2}{8\pi}$ and the turbulent kinetic energy density $\rho \epsilon_{turb}$ in this condition \citep{Subramanian_1998MNRAS}. A detailed description and the physical verification of the model can be found in \citet{Paul_2018arXiv}.

It is well known that the thermally distributed ICM particles, after shock acceleration (through DSA), or turbulent re-acceleration (TRA), produce a power-law energy distribution 
\begin{equation}
\left(\frac{dn_e}{dE_e} \right) \propto E^{-\delta}
\end{equation}
where $\delta$ is the spectral index of the injected electron energy. For DSA, the value is 2 or more steeper 
\citep{Drury_1983RPPh}. For TRA, the electron energy power-law with a simple assumption would be to consider a fully developed Kolmogorov type turbulence i.e. $\delta$ to be $\frac{5}{2}$ \citep{Gouveia_2005A&A,Fang_2016JCAP}. 

Synchrotron radio emission is thus calculated using the standard synchrotron emission formula 
\begin{eqnarray}
 \frac{d^{2}P(\nu_{obs})}{dVd\nu} &=& \frac{\sqrt{3}e^3B}{8m_e c^2}\,\int_{E_{\rm min} }^{E_{\rm max}}dE_e\,F\left(\frac{\nu_{obs}}{\nu_c}\right)\,\left(\frac{dn_e}{dE_e} \right)_{\rm inj}
\end{eqnarray}
where $F(x)=x\int_x^\infty K_{5/3}(x')dx'$ is the synchrotron function, $K_{5/3}$ is the modified Bessel function, and $\nu_c$ is the critical frequency of synchrotron emission, 
\begin{equation}
\nu_c =  {3}\gamma^2{eB}/{(4\pi m_e)} = 1.6\, ( {B}/{1 \mu \rm G})({E_e}/{10 \rm GeV})^2 ~ \rm GHz 
\end{equation}
$\frac{dn_e}{dE_e}$ is the injected electron energy spectrum determined either by DSA \citep{Hong_2015ApJ} or the TRA mechanism \citep{Fang_2016JCAP}. Here, we did not include synchrotron aging model, which may lead to a slightly different spectrum of radio emission at low frequencies. For a detailed account of the computation of radio emissions for this study, see \citet{Paul_2018arXiv}.

A cluster merger scenario from our cosmological simulation is plotted in Figure~\ref{fig:radio-emission}. The columns from left to right represent specific phases of the galaxy cluster merger stage. The top row shows density maps. Figure~\ref{fig:radio-emission}(a) clearly show multiple groups approaching each other i.e. pre-merger state. Figure~\ref{fig:radio-emission}(b) shows that the groups merged into a single structure. Finally, Figure~\ref{fig:radio-emission}(c) shows that the structure becomes almost spheroidal during the post merger relaxation phase. Further rows are for the same states as the above but representing temperature, vorticity magnitude ($\bar{\omega}=\nabla\times\bar{\rm{v}}$, representing turbulence) and finally computed radio map, from top to bottom respectively.

The temperature maps show, at pre-merger state, there is only a weak compression shock (most whitish part in the map). Merger shock launched at the centre of the panel~(e) and after about 1.2 Gyr of traveling through the cluster medium, it crosses little more than an Mpc (see panel~(f)) to reach the cluster periphery, similar to the observed cluster Abell 1697, though two sided. In the vorticity map in panel~(i), it can be noticed that the turbulence is generated behind the shock front as well as within a long trail of wake turbulence almost reaching the cluster centre (encircled by white dashed line), specifically for the shock on the right side of the image. The computed radio emission for the same (in panel~(l)) shows a long radio emission trail (encircled by black dashed line) behind the shock (simulated radio relic), generated due to turbulent re-acceleration from the wake turbulence.

\end{appendix}

\end{document}